\documentclass[onecolumn,12pt,nofootinbib,prd,floatfix,preprintnumbers,amsmath,amssymb,groupedaddress,superscriptaddress]{revtex4}
\usepackage{amsmath}
\usepackage{verbatim}
\usepackage{enumerate}
\usepackage{amsthm} 
\usepackage{xcolor}
\usepackage{stmaryrd}
\usepackage{tikz-cd}

\newtheorem{thm?}[thm]{Theorem?}
\theoremstyle{remark}

\theoremstyle{definition}

\newcommand{\Cat}{\mathsf{Cat}}

\newcommand{\Fun}{\mathsf{Fun}}
\newcommand{\Map}{\mathrm{Map}} 
 
\newcommand{\bbC}{\mathbb{C}} 
\newcommand{\bbN}{\mathbb{N}} 
\newcommand{\bbR}{\mathbb{R}} 
\newcommand{\bbZ}{\mathbb{Z}}
\newcommand{\Bord}{\mathsf{Bord}}
\newcommand{\Alg}{\mathsf{Alg}} 
\newcommand{\Vect}{\mathsf{Vect}} 
\newcommand{\Aut}{\mathrm{Aut}}
\newcommand{\End}{\mathrm{End}} 
 
\newcommand{\TFT}{\mathrm{TFT}}
\newcommand{\gTFT}{\Theta}

\begin{document}   
\title{Generalized symmetries of topological field theories} 
\date{\today}
\author{Ben Gripaios}  
\email{gripaios@hep.phy.cam.ac.uk}
\affiliation{
Cavendish Laboratory,
JJ Thomson Avenue,
Cambridge, UK}
\author{Oscar Randal-Williams}  
\email{o.randal-williams@dpmms.cam.ac.uk}
\affiliation{
  DPMMS, University of Cambridge, Wilberforce Road, Cambridge, UK}
\author{Joseph Tooby-Smith}  
\email{j.tooby-smith@cornell.edu}
\affiliation{Department of Physics, LEPP, Cornell University, Ithaca, NY 14853, USA}
%%%%%%%%%%%%%%%%%%%%%%%%%%%%%%%%%%%%%%
\begin{abstract} 
  %%%%%%%%%%%%%%%%%%%%%%%%%%%%%%%%%%%%%%
We study generalized symmetries in a simplified arena in which the
usual quantum field theories of physics are replaced with topological
field theories and the smooth structure with which the symmetry groups
of physics are usually endowed is forgotten. Doing so allows many
questions of physical interest to be answered using the tools of
homotopy theory. We study both global and gauge symmetries, as well as
`t~Hooft anomalies, which we show fall into one of two classes. Our approach also allows some insight
into earlier work on symmetries (generalized or not) of topological
field theories.
%%%%%%%%%%%%%%%%%%%%%%%%%%%%%%%%%%%%%% 
\end{abstract}   
%%%%%%%%%%%%%%%%%%%%%%%%%%%%%%%%%%%%%%
\maketitle
\tableofcontents
%%%%%%%%%%%%%%%%%%%%%%%%%%%%%%%%%%%%%%
%%%%%%%%%%%%%%%%%%%%%%%%%%%%%%%%%%%%%%
\section{Introduction} 
  %%%%%%%%%%%%%%%%%%%%%%%%%%%%%%%%%%%%%%
Generalized symmetries have come to play an important r\^{o}le in
quantum field theory. Nevertheless, they retain an air of mystery.
In
\cite{Gaiotto:2014kfa}, a $q$-form global symmetry in spacetime dimension $d$ was defined as
`topological operators $U_g(M^{(d-q-1)})$ associated to
codimension $q+1$ manifolds $M^{d-q-1}$ that fuse according to a group
law
$U_g (M^{(d-q-1)}) U_{g^\prime} (M^{(d-q-1)}) = U_{g^{\prime
    \prime}} (M^{(d-q-1)})$' (where $g^{\prime \prime}= g
g^\prime$), but this leaves us with several questions. Firstly, the definition makes sense for every $0\leq q+1
\leq d$, so do the individual groups assemble themselves into some
larger
structure, and if
so how? Secondly, we
know that even in quantum mechanics, there exist ordinary symmetries in which the associated operators
do not quite fuse according to a group law (or to put it more prosaically,
do not form a representation). For example, a rotation of an electron is
represented projectively and a reversal of time is
represented antilinearly.
Can such symmetries be
generalized?
Thirdly, what does `topological' mean, exactly, and how does it
relate to our usual understanding of ordinary symmetries in quantum mechanics as
(unitary) operators that commute with the hamiltonian operator? Fourthly,
can we give a meaning to gauging a generalized symmetry? If so, can it always be
done, or are there possible 't~Hooft anomalies? If it can be done, is the
resulting theory unique?

Here, we wish to shed some light on such questions by studying
generalized symmetries in a simplified arena in which we replace the usual dynamical quantum field
theories of physics with topological field theories and we forget the
smooth Lie group structure with which the symmetry groups of physics are usually
endowed (though we make some remarks about Lie groups and the
connection to Noether's theorem in \S \ref{sec:lie}).

Such theories are somewhat boring,
dynamically speaking, in that the theories are few and far between and, in
any given one of them, very little can actually happen. But, as we will see, this
  disadvantage is 
  offset by the consequent advantage that they exhibit
  larger amounts of symmetry than theories with additional structures. Moreover, they can be formulated precisely
using the language of category theory, and we can often calculate
everything we desire. (Indeed, many of the mathematical constructions
we describe are known to mathematicians, or at least will come as no surprise, but the
interpretation in terms of generalized symmetries of physics is hopefully new.)

Generalized symmetries by
their very nature require us to use higher categories, which in
practice involves a great deal of faff. Fortunately, almost all of it can be avoided by
going all the way to infinity and observing that 
both the generalized symmetries and the
topological field theories on which they act form very special cases of
$\infty$-categories, namely $\infty$-groupoids, in which all morphisms are invertible. 
As such, they can be replaced by topological spaces, or rather homotopy types, and
the requisite mathematics can mostly be phrased in terms of
homotopy theory. To translate back to physics requires us to return to
higher categories, whose terminology we use in a colloquial
sense, except when it comes to concrete examples. 

So, for
example, a point $Z$ in a topological space $\gTFT$ representing some
$\infty$-groupoid of topological field theories corresponds to an
object, {\em i.e.} a specific
topological field theory, a path
between two points corresponds to a 1-morphism between the
corresponding topological field theories, a homotopy between two
paths corresponds to a 2-morphism, and so on. 

Our questions above are easily answered using this
language. To give a sketch,
it is convenient to distinguish the three mathematical notions
of a {\em group}, an {\em action} of that group, and a
{\em fixed point} of
that action, and to generalize each of these.

In our simplified arena, a generalized group can be
completely characterized by the homotopy type of
a pointed connected topological space, or equivalently the classifying space
$BG$ of a single
{\em topological} group $G$ (which is not unique). The semi-infinite tower of homotopy groups
$\pi_{q+1}(BG) \simeq \pi_q(G)$ encode the abstract groups
of $q$-form symmetries for each non-negative integer $q$.
The tower comes equipped with a rich structure. For example, for
$q\geq 1$ the groups are abelian, in agreement with expectations for
generalized symmetries. Furthermore, 
there is an
action of the ordinary symmetry group $\pi_0(G)$ on each of the generalized
symmetry groups $\pi_q(G)$, induced by the action of the topological
group $G$ on itself by
conjugation.
Thus, an answer to our first
question is: a generalized group is the classifying space of a
topological group (and an
ordinary group corresponds to the special case of a discrete group). 

A generalized action of a generalized group on some space $\gTFT$ of topological
field theories is then a fibration over $BG$, together with an
identification of  $\gTFT$ with the fibre over the basepoint.  (Many,
but not all, of these arise via a continuous action of $G$ on $\gTFT$
as the bundle $EG \times_G
\gTFT \to BG$.) A generalized fixed point is a homotopy fixed point,
namely a section of the
fibration (in the case of the bundle $EG \times_G
\gTFT \to BG$, this is equivalently a $G$-equivariant
map from $EG$ to $\gTFT$).

Our second question can then be answered as follows. Every
space $\gTFT$ admits the trivial action of $G$, in which no points are
moved. By passing back to the language of category theory, one sees
that the corresponding homotopy fixed points (which are simply maps
$BG \to \gTFT$) correspond to
true representations. But a specific $\gTFT$ may also admit non-trivial
actions of $G$, and we will see in examples how these reproduce projective
and antilinear representations, and more besides.
In
category-theoretic language, these exotic possibilities arise
because topological
field theories (and presumably quantum field theories in general)
form an $\infty$-groupoid, whose morphisms record (some of) their internal
structure. So the right notion of a (generalized) group action is
  not one on a set, but on an $\infty$-groupoid, and the right notion
  of a fixed point is a limit in the sense of
  $\infty$-categories.

To answer the third question, consider again the homotopy fixed points
of the trivial action, {\em i.e.} the maps $BG \to \gTFT$. Looping such a map, we
obtain a map from $G$ to the space, $\Omega_Z
\gTFT$, of loops in $\gTFT$ based at $Z \in \gTFT$. On the category-theoretic side, these correspond to the
automorphisms of the topological field theory $Z$, whose objects are
invertible natural transformations from the topological field theory
to itself. They thus commute 
with all possible
dynamical evolutions. This is just the same as what happens for
ordinary symmetries in quantum mechanics, except that in topoogical
field theories the possible dynamical evolutions are different in
nature, being evolutions along spacetimes with non-trivial topology or
along spacetimes equipped 
with non-trivial geometric structures. There are induced homomorphisms
$\pi_q(G) \to \pi_{q+1}(\gTFT,Z)$, showing that the composition of natural
transformations respects the group law on the nose. For non-trivial
actions, we get transformations whose composition does not respect the
group law on the nose, but is merely coherent with respect to
it.

On the category theory side, a $q$-form symmetry corresponds to choices of $(q+1)$-transfors of TFTs (where a $0$-transfor is a functor, a $1$-transfor a natural transformation, etc.). Part of the data of such a $(q+1)$-transfor is the assignment of top-level morphisms in the target category to codimension $(q+1)$-manifolds. This is in accordance with the notion of a $q$-form symmetry given in~\cite{Gaiotto:2014kfa}. In particular, $0$-form symmetries assign  top-level morphisms to codimension $1$-manifolds. On looping, these morphisms descend down to linear maps between vector spaces.

To deal with the fourth question, of gauging generalized symmetries, we equip
the bordism category underlying topological field theories with
tangential structure. One can do this in a very general way, following
Lurie \cite{Lurie:2009keu}, that allows for gauge symmetries that act non-trivially
on spacetime. By introducing a notion of fibrations of tangential structures,
we construct maps, in the language of
homotopy theory, which we call {\em globalization maps}, that
send spaces of theories with gauge symmetry to spaces of theories with
global symmetry. 
This gives a convenient framework for discussing `t~Hooft anomalies,
which can be seen to be of one of two kinds. The first is an anomaly
afflicting an entire space of theories with global symmetry, so we
call it a {\em metaphysical 't~Hooft
    anomaly}, and
arises when that space is not the image of any
globalization map. 
An example familiar from quantum mechanics are theories with genuinely
    projective representations (meaning the associated 2-cocycle is
    not a coboundary) of an ordinary symmetry group.

    The second kind of anomaly
    is that even if a suitable globalization map exists, it may fail
    to be
    surjective (on $\pi_0$); a theory with a global symmetry lying outside the
    image will then be anomalous. We call these {\em unphysical `t~Hooft
    anomalies}, for reasons which will soon become clear.

    These considerations also show that one can have what we call {\em`t
    Hooft ambiguities}. Namely, even if a theory is non-anomalous, so
    it is in the image of some globalization map, there is no guarantee
    that that map is unique, nor that it injects. So there may be many
    ways to gauge a global symmetry.

    In the case of
    topological field theories that are fully local (or fully extended in
    the mathematicians' jargon), the cobordism hypothesis implies that
    the globalization maps are homotopy equivalences, so
    unphysical anomalies and ambiguities are necessarily absent. 
    They thus
    arise purely in theories that fail to fully respect the sacred
    physics principle
    of locality, hence the moniker unphysical.

    Our approach also allows us to shed further light on several earlier
    observations in the literature regarding symmetries
    (generalized or ordinary) of topological field theories.

    To set the scene for this, consider the example of the
    orientable topological field theory in
    $d=2$ obtained by quantizing a classical field theory with gauge
    symmetry  $\bbZ_n^2$, as described
    in \cite{Dijkgraaf:1989pz}. For now we will be deliberately vague regarding whether
    this theory is considered to be fully extended or not, as well as
    regarding what
    the target category is. 
    
The classical action of this theory is specified by $p \in 
   H^2(B(\bbZ_n^2), \bbC^\ast) \simeq \bbZ_n$.  In
   \cite{Gaiotto:2014kfa}, it was argued that this
   theory has $0$- and $1$-form symmetries
   given by $\bbZ_{n/k}^2$, where $k =
   \mathrm{gcd}(p,n)$.
To make sense of this in our language requires us to consider the theories to be fully
extended (with values in a certain bicategory of algebras over $\bbC$). 
  We will then show that in fact  $\bbZ_{n/k}^2$ is merely a subgroup of $\pi_1(\gTFT,Z)
   \simeq S_{n^2/k^2}$ ({\em i.e.} the permutation group on $n^2/k^2$ elements) and of $\pi_2(\gTFT,Z)
   \simeq \left({\bbC}^\ast\right)^{n^2/k^2}$ for the corresponding
   theory $Z$. These larger symmetries cannot be seen by inspection of
   the classical action. More generally, we
   show how to compute $\pi_1(\gTFT,Z)$ and $\pi_2(\gTFT,Z)$ for every
   fully-extended topological field theory in $d=2$ and show how they arise as
   subgroups of $S_m$ and $\left({\Bbbk}^\ast\right)^{m}$, for some
   $m \in \bbN$ (which are $\pi_1(\gTFT,Z)$ and $\pi_2(\gTFT,Z)$ for every
   possible $Z$ for the base
   case of topological field theories defined on manifolds equipped
   with 2-framings) with values in algebras over any separably
     closed field $\Bbbk$. We also show how one may characterize all possible homotopy fixed
   points of all possible actions in such cases.

These results are suspiciously close to those of \cite{Gukov:2021swm},
which showed that 0-form symmetries of
   {\em un}extended oriented topological field theories whose corresponding
   commutative Frobenius algebras are semisimple act 
  on a basis of idempotents by permutations preserving the trace map, while 1-form symmetries act
  by multiplication of idempotents by elements of $\bbC^\ast$. In fact this is no
  coincidence, because the semisimple commutative Frobenius algebras
  correspond precisely to oriented topological field theories which
  are extend{\em able}: by extending, one can make a proper definition
  of a 1-form symmetry, and show that the conditions obtained in
  \cite{Gukov:2021swm} are not only necessary, but also sufficient. 
  Our results
   thus not only generalize those of \cite{Gukov:2021swm}, but also place them in
   their proper context.

   The same (unextended) $\bbZ_n^2$ gauge theory appears
   elsewhere \cite{Kapustin:2014zva}  as an example of a theory with ordinary global symmetries
   that ostensibly
   suffer from a `t~Hooft anomaly. This seems odd, given that the
   theory is extendable, and given that for extended theories the
   cobordism hypothesis implies that the
   globalization map is a homotopy equivalence. The
   resolution of this apparent paradox is as
   follows. Ref.~\cite{Kapustin:2014zva} in fact describes a
   method for constructing theories with global symmetries that are
   free of `t~Hooft anomalies (see the
   diagram in Eq. \ref{eq:Mueller}). The construction only works if a
   certain necessary condition is
   satisfied. Ref.~\cite{Kapustin:2014zva} defines a theory to be `anomalous' if that
   condition is violated, but that merely means that the construction
   cannot be carried out. As a result one does not even have a theory
   of which one can ask the question of whether it is anomalous or
   not, in the usual sense.  

   The outline of the paper is as follows. In \S\ref{sec:tft}, we give a brief
   introduction to $\infty$-categories and describe some examples relevant for
   topological field theory. In \S\ref{sec:glo} we define generalized
   groups and generalized global symmetries of topological field
   theories. In \S\ref{sec:gau}  we define generalized gauge
   symmetries, construct the globalization maps, and define 't~Hooft
   anomalies and ambiguities in that context. In \S\S \ref{sec:egunext}--\ref{sec:egnmext} we discuss examples of
   topological field theories in $d=1$ and $d=2$. In \S\ref{sec:lie}
   we discuss Lie group symmetries and in \S\ref{sec:disc} we compare our
   results with earlier literature.
%%%%%%%%%%%%%%%%%%%%%%%%%%%%%%%%%%%%%%
\section{Topological field theories\label{sec:tft}} 
%%%%%%%%%%%%%%%%%%%%%%%%%%%%%%%%%%%%%%
    %%%%%%%%%%%%%%%%%%%%%%%%%%%%%%%%%%%%%%
    \subsection{Unextended topological field theories\label{sec:unextft}}
    %%%%%%%%%%%%%%%%%%%%%%%%%%%%%%%%%%%%%%
     To set the scene, we begin with a review of unextended topological
   field theories, formulated using the category-theoretic approach
   pioneered by Atiyah, Kontsevich, and Segal.
We need the notions of category, functor, natural transformation, and
equivalence of categories, all of which are standard and may be found
in \cite{maclane:71}. To set the notation, a category\footnote{A
    note on notation: since categories and higher categories can be
    thought of as special cases of infinity categories, we use a sans
    serif font for all of them (and even the same letter $\mathsf{C}$
    to denote a generic one), except in the case of
    $\infty$-groupoids, which will later be treated as topological
    spaces.}
$\mathsf{C}$ is a collection
of objects with a
set of composable morphisms between each pair of objects; given a pair of
categories $\mathsf{C}$ and $\mathsf{D}$, the functors between them themselves form the
objects of a category $\Fun(\mathsf{C},\mathsf{D})$ whose morphisms are the natural
transformations. 

We also need extra structure on a category $\mathsf{C}$,
 namely a symmetric monoidal structure, for which definitions may be found
 in \cite{etingof_tensor_2016}. Roughly, we have a functor $\otimes: \mathsf{C}
 \times \mathsf{C} \to \mathsf{C}$, a unit object $1 \in \mathsf{C}$, natural braiding isomorphisms $s_{a,b} : a \otimes b \overset{\sim}\to b \otimes a$ that square to the identity, and natural isomorphisms $a \otimes 1 \overset{\sim}\to a$ exhibiting $1$ as a unit. 
A dual to $a$ is an object $a^\vee$ along with
 an evaluation morphism $\mathrm{ev}:a^\vee \otimes a \to 1$ and a coevaluation
 morphism $\mathrm{coev}:1 \to a \otimes a^\vee$
 obeying certain familiar conditions; we say that $a$ is invertible if these morphisms
 are, moreover, isomorphisms. Given a pair of symmetric monoidal
 categories, there is a notion of a (strong) symmetric monoidal functor; these
 form the objects of a symmetric monoidal category $\Fun^\otimes(\mathsf{C},\mathsf{D})$,
 whose morphisms are
 the monoidal natural transformations.

 For unextended topological field theory in spacetime dimension $d$,
 we start from the symmetric monoidal category $\Bord_{d,1}$. An
 object in $\Bord_{d,1}$ is a closed ({\em i.e.} compact without
 boundary) $(d-1)$-manifold $M$. (In
 general, we may wish to equip manifolds with additional structures,
 such as an orientation or spin structure, but
 since this will not play a significant r\^{o}le until \S~\ref{sec:gau}, we
elide it for now.) A morphism from $M$ to $N$ in $\Bord_{d,1}$ is an
 equivalence class of
 compact $d$-manifolds $W$ whose boundary is identified with the disjoint union
 $M \coprod N$, where two $W$'s
 are considered equivalent if they are related by a
 diffeomorphism
 which is the identity on the boundary.\footnote{We remark that this unpleasantry is
   already a good motivation to go to $(\infty,1)$-categories, in which a
   morphism is simply a bordism.} Composition of bordisms is defined
 by gluing manifolds along the appropriate boundary components. The product in the symmetric monoidal
 structure is given by the disjoint union of manifolds (so we denote
 it $\coprod$) and the unit is
 given by the empty manifold $\emptyset$. Physically, $W$ represents the
 `spacetime' of a euclidean quantum field theory evolving in euclidean
 time from space
 $M$ to space $N$, but the evolution is allowed to be topologically
 non-trivial.  

 An unextended topological field theory is then a symmetric monoidal
 functor $Z$ out of $\Bord_{d,1}$ to some target symmetric monoidal
 category, which we must now choose. Given that we are modelling a
 euclidean theory, and given that there is 
is no obvious
 available notion of a Wick rotation to a lorentzian theory, it seems
 artificial to demand the usual quantum-mechanical structure of
 complex Hilbert space and operators that are self-adjoint or isometries (or
 hermitian and unitary in the physics lingo), or even some Wick-rotated
 version thereof. We therefore choose the target to be 
 $\Vect_{\Bbbk}$ whose objects are vector spaces over an arbitrary
 field $\Bbbk$, whose morphisms are
 $\Bbbk$-linear maps, and whose product is the usual tensor product of
 vector spaces (so the unit may be taken to be $\Bbbk$). Physically,
 the fact that $Z$ is a functor, so preserves composition, encodes (partially)
 the fact that theories of physics should be local. Indeed, going
 backwards we see that we can recover the evolution along $W$ from the
 evolution along bordisms obtained by cutting it along an arbitrary 
submanifold of
 codimension one.\footnote{We remark that the fact that we are only
   allowed to cut along codimension one means that, locality is not
   fully manifest. This is
   already a good motivation to go to $(\infty,d)$-categories.} The
 symmetric monoidal structure of $Z$ allows for composite systems to
 be entangled in the usual way.

 The correlation functions of the
 theory are encoded as follows. The usual local operators of
 quantum field theory are supported on points, and the effect of
 inserting such operators at points $w_1,\dots,w_i,\dots,w_n$ in a closed $d$-manifold $W$ is found by deleting
 disjoint open neighbourhoods of each $w_i$ in $W$, resulting in a
 compact $d$-manifold with boundary $W^\prime$, which in turn defines
 a bordism from $\coprod_i S^{d-1}$ to the empty $d$-manifold
 $\emptyset$. Applying the functor $Z$ returns a linear map $Z(W^\prime):
 \bigotimes_i Z(S^{d-1}) \to \Bbbk$. The vectors in $Z(S^{d-1})$ may thus
 be regarded as the local operators of the theory, and the linear map
 $Z(W^\prime)$, which returns a complex number given a choice of local
 operator at each $w_i$, may be regarded as the correlation function. 

 Key to our story will be the category $\TFT_{d,1}
 := \Fun^\otimes(\Bord_{d,1},\Vect_{\Bbbk})$ of symmetric monoidal functors. Its objects are topological field
 theories and its morphisms (which are monoidal natural
 transformations) give us a way to compare topological field theories
 with one another and thus detect at least some of their structure. In
 particular, the presence of a morphism between two theories that is
 an isomorphism allows us to conclude that they are 
are physically
 equivalent, since they will
 lead to theories in which the correlation functions (and ultimately
 the observables)
 are related to one another in the same way.

 In fact, every morphism in $\TFT_{d,1}$ is a isomorphism (so
 $\TFT_{d,1}$ is a groupoid). 
To show this, we need to show that given a natural
 transformation $\eta: Z \to Z^\prime$ and any closed
 $d$-manifold $M$, the induced linear map $\eta_M : Z(M) \to
 Z^\prime(M)$ is an isomorphism. Regarding the `cylinder'  $M\times I$
 as a bordism $M \coprod M \to \emptyset$ or $\emptyset \to M \coprod M$
 and applying $Z$ furnishes us with evaluation and coevaluation maps
 that exhibit $Z(M)$ as a dual of itself and the dual map $\eta_M^\vee: Z^\prime(M)
 \to Z(M)$ turns out to be the sought-after inverse to $\eta_M$.

Even though every morphism in a groupoid such as $\TFT_{d,1}$ is an
 isomorphism, the groupoid can tell us much more than just whether two
 theories are equivalent, because each theory (and indeed any object
 in any category) has associated to it a group of automorphisms,
 namely the isomorphisms from the theory to itself.
It is natural to guess that this group is related to the global
symmetry of the theory and this guess is confirmed by picking apart
the definition of an automorphism of $Z \in \TFT_{d,1}$: it is, for each closed $d$-manifold $M$, a linear isomorphism
$\eta_M:Z(M) \to Z(M)$ such that, for any bordism $W: M \to N$, the
diagram
\begin{equation}
\begin{tikzcd}
Z(M) \arrow{d}{\eta_M}\arrow{r}{Z(W)}& Z(N)\arrow{d}{\eta_N}\\
Z(M) \arrow{r}{Z(W)}& Z(N)
\end{tikzcd}
\end{equation}
commutes. So the components of $\eta$ are linear maps (for each state
space $Z(M)$) that commute
with the dynamical evolutions $Z(W)$ along all possible euclidean
spacetimes $W$. This looks very close to the usual quantum-mechanical notion of a unitary
operator on the Hilbert space of states that commutes with the unitary
time evolution operator, except that the notion of unitarity has gone
and that the evolutions are now trivial (since the cylinder $M\times
I$ is the identity bordism on $M$, it gets sent by the functor $Z$ to the identity linear map on $Z(M)$), unless spacetime is
topologically non-trivial. Moreover, the diagram shows that
the components of $\eta$ are compatible with locality, expressed in terms of cutting and pasting
of bordisms.

We could, therefore, make an intrinsic definition of the global
symmetry group of $Z$
to be
its group $\Aut(Z)$ of monoidal natural automorphisms, or alternatively make an
extrinsic definition of a global symmetry of $Z$ as a group $G$
together with a homomorphism $G \to \Aut(Z)$, but we will see in the
next Section that it pays to do something which is na\"{\i}vely rather different, namely to
consider fixed points (in a appropriate sense) of actions of $G$ on
the groupoid $\TFT_{d,1}$. In fact this turns out to generalize the
notion of a homomorphism $G \to \Aut(Z)$ (which is recovered as
a fixed point of the trivial action of $G$ on $\TFT_{d,1}$). Doing so allows us to
capture the notion that an element of a physical symmetry group
need not  fix $Z$, but rather can send it to an isomorphic theory,
without affecting physical observables. We will see in \S \ref{sec:egmext},
moreover, that this generalization is needed to describe well-known
physical phenomena such as the behaviour of electrons under spatial
rotations and time-reversal invariance.

Before doing that, we describe extended topological field theories.
These will be needed not only to define generalized global symmetries in
\S\ref{sec:glo}, but also to formulate physics in a way which is fully
local. As we will argue in \S  \ref{sec:egmext}, certain `t~Hooft anomalies are best viewed as arising
    from a failure to define a theory in such a way and so should be
    regarded as unphysical. 
%%%%%%%%%%%%%%%%%%%%%%%%%%%%%%%%%%%%%%
    \subsection{Extending topological field theories downwards\label{sec:extft}}
    %%%%%%%%%%%%%%%%%%%%%%%%%%%%%%%%%%%%%%
    We have already seen that defining topological field theories
    using ordinary categories only captures a part of the local structure of
    physics. To fully capture locality, the theory should be defined
    not only on closed $(d-1)$-manifolds and $d$-manifolds with
    boundary, but on manifolds with corners of all possible
    codimensions, so that dynamics can be reconstructed by pasting
    together simplices.

For that, we require higher categories. Roughly, these should consist of objects,
morphisms, higher morphisms, and so on, which can be composed in multiple
ways in a coherent fashion. Precise definitions are, however,
somewhat involved. Since we will only go one step higher in
our examples, and since we will anyway soon need the yet more general
notion of an $\infty$-category, we will content ourselves here with sketching the simplest case,
namely a bicategory. Full details are given in, {\em e.g.}, \cite{schommer2009classification}.
  
A \emph{bicategory}, $\mathsf{C}$, is a collection of objects with a category $\mathsf{C}(a,b)$
for each ordered pair $(a,b)$ of objects in $\mathsf{C}$. The objects of
$\mathsf{C}(a,b)$ are called \emph{1-morphisms} and the morphisms of $\mathsf{C}(a,b)$ are called
\emph{2-morphisms}. In addition, there is a functor $\mathsf{C}(b,c)\times
\mathsf{C}(a,b)\rightarrow \mathsf{C}(a,c)$ known as \emph{horizontal composition}, with
a unit 1-morphism $1_a \in \mathsf{C}(a,a)$
whilst composition within $\mathsf{C}(a,b)$ is called \emph{vertical
  composition}. An \emph{equivalence} between objects $a$ and $b$ is a
pair of 1-morphisms $f: a \leftrightarrow b: g$ and a pair of
2-morphisms $\alpha: 1_a \to g \circ f$ and $\beta: f \circ g \to
1_b$\footnote{We reluctantly perpetuate the now-standard practice of denoting both horizontal
composition of 1-morphisms and vertical composition of 2-morphisms
with $\circ$, with $*$ being used for horizontal composition of
2-morphisms.} that are isomorphisms in $\mathsf{C}(a,a)$ and $\mathsf{C}(b,b)$
respectively.
      
 Given two bicategories, we have the notion of a functor
    between them; given two functors we have the notion of a
    transformation, and given two transformations we have the notion
    of a modification. The functors, transformations, and
    modifications assemble themselves respectively into the objects, 1-morphisms,
    and 2-morphisms of a functor bicategory.

  We will also need a symmetric monoidal structure on
    bicategories and the corresponding bicategory of symmetric
    monoidal functors.

 An example of a symmetric monoidal bicategory, which will play the r\^{o}le of the
    target bicategory in our examples, is $\Alg_\Bbbk$: an object
    is an algebra over a field $\Bbbk$, a 1-morphism from an algebra $A$
    to an algebra $B$ is an $(A,B)$-bimodule, and a 2-morphism is an
    $(A,B)$-bilinear map. The horizontal composition of 1-morphisms is
    given by the tensor product of bimodules (over the algebra in the
    middle) and the symmetric monoidal structure is given by the
  tensor product over $\Bbbk$. The relevance of this bicategory to
  physics is as follows. Given any (symmetric) monoidal bicategory
  $\mathsf{C}$, the endomorphisms of the unit object form a
  (symmetric) monoidal category, which we denote $\Omega
  \mathsf{C}$. This looping construction extends to higher monoidal
  categories and is adjoint to a delooping construction, which sends a
  monoidal higher category $\mathsf{C}$ to a monoidal category
  $B\mathsf{C}$ one level higher with a single object whose
  endomorphisms are $\mathsf{C}$. The unit object of the bicategory
  $\Alg_\Bbbk$ is $\Bbbk$ and its endomorphism category consists of
  $\Bbbk$-vector spaces and $\Bbbk$-linear maps. Thus, $\Alg_\Bbbk$
  may be regarded as an extension (not unique) of the usual target
  category $\Vect_\Bbbk$ of unextended topological field
  theories. Looping again, we obtain the symmetric monoidal 0-category
  (\emph{i.e.} commutative monoid) of linear endomorphisms of $\Bbbk$, which
  is isomorphic to $\Bbbk$ itself. This provides a target for
  maximally {\em un}extended theories, in which the source `category'
  contains only closed $d$-manifolds, which we discuss as a toy
  example in \S \ref{sec:egunext}.
    %%%%%%%%%%%%%%%%%%%%%%%%%%%%%%%%%%%%%%
    \subsection{Extending topological field theories upwards\label{sec:infextft}}
    %%%%%%%%%%%%%%%%%%%%%%%%%%%%%%%%%%%%%%
    It is convenient, for a number of reasons, to extend topological
    field theories upwards as well, using the language of
    $\infty$-categories. One is that, as we have already hinted, it
    leads to a simplification of the domain. 
Another is that the connection with homotopy theory is more explicit. A third is that it then becomes obvious that symmetries of topological field theories should be described by homotopy fixed points of group actions, since homotopy limits are the only kind of limits 
    in the $\infty$-categorical context.

As for higher categories, we shall content ourselves with a sketch of
the relevant concepts and definitions. For more details, see
\cite{Lurie:2009keu}.

An $(\infty,n)$-category $\mathsf{C}$ has objects, and morphisms of all levels, where the morphisms at level greater than $n$ are invertible, but now in a recursive sense. So a morphism $f$ is invertible if there exists a morphism $g$ in the other direction along with morphisms $\alpha:1 \to g\circ f$ and $\beta: f \circ g \to 1$ at one level higher that are themselves invertible. An $\infty$-category is $n$-truncated if the morphisms at level greater than $n$ are equivalent to identity morphisms. Evidently, a higher $n$-category can be identified with an $n$-truncated $\infty$-category (which we hope excuses the somewhat overloaded notation). Going in the other direction, we obtain the homotopy $n$-category of an $\infty$-category $\mathsf{C}$ by replacing the morphisms at level $n$ with their equivalence classes under the equivalence described above.

Given two $(\infty,n)$-categories $\mathsf{C},\mathsf{D}$, there is an
$(\infty,n)$-category of functors $\Fun(\mathsf{C},\mathsf{D})$ from
one to the other, and if $\mathsf{D}$ is $n$-truncated then $\Fun(\mathsf{C},\mathsf{D})$ is an $n$-truncated category. There is an $(\infty,n+1)$-category $\Cat_n$, whose objects are $(\infty,n)$-categories and whose endomorphisms are the $(\infty,n)$-categories of functors.

An $(\infty,0)$-category, in which all morphisms are invertible, is also called an $\infty$-groupoid and corresponds, via the homotopy hypothesis, to a homotopy type, {\em i.e.} a topological space up to weak homotopy type. In one direction, this correspondence is given by forming the fundamental $\infty$-groupoid of a topological space $X$: objects are points in $X$, 1-morphisms are continuous paths, 2-morphisms are homotopies between paths, and so on. 

The $(\infty,p)$-category $\Bord_{d,p}$ has objects given by $(d-p)$-manifolds (with suitable corners), 1-morphisms given by $(d-p-1)$-manifolds,$ \dots$, $p$-morphisms given by $d$-manifolds, $(p+1)$-morphisms given by diffeomorphisms of $d$-manifolds, $(p+2)$-morphisms given by isotopies of diffeomorphisms, and so on. 

In this picture, topological field theories are $(\infty,p)$-functors from $\Bord_{d,p}$ to some target $(\infty,p)$-category $\mathsf{D}$. An argument similar to the one we gave for unextended topological field theories shows that the category 
$\Fun^\otimes(\Bord_{d,p},\mathsf{D})$ is in fact an $\infty$-groupoid, or a homotopy type~\cite[Remark 2.4.7]{Lurie:2009keu}. In our examples, we will take $\mathsf{D}$ to be $p$-truncated, so that a topological field theory factors through the homotopy $p$-category of $\Bord_{d,p}$. Moreover, $\Fun^\otimes(\Bord_{d,p},\mathsf{D})$ is a homotopy $p$-type. 
%%%%%%%%%%%%%%%%%%%%%%%%%%%%%%%%%%%%%%
\section{(Generalized) global symmetries\label{sec:glo}} 
  %%%%%%%%%%%%%%%%%%%%%%%%%%%%%%%%%%%%%%
 %%%%%%%%%%%%%%%%%%%%%%%%%%%%%%%%%%%%%%
    \subsection{Ordinary global symmetries\label{sec:oglo}}
    %%%%%%%%%%%%%%%%%%%%%%%%%%%%%%%%%%%%%%
    Before discussing generalized global symmetries, let us consider
    how to describe ordinary global symmetries of topological field
    theories. As usual in mathematics, it is convenient to separate
    the notions of a group, an action of that group on something, and a fixed point
    of that action.

    The most straightforward example is that of a group action on a
    set $S$, where a group action is a homomorphism from $G$ to the
    group $\mathrm{Aut}(S)$ of bijections of $S$. A fixed point is an
    element of $S$ that is fixed by each element of $G$ and the set of
    all fixed points forms a subset $S^G$ of $S$.

    This can be formulated using the language of category theory as
    follows. A group $G$ can be considered as a category (in fact,
      a groupoid) $BG$
    with a single
    object and an isomorphism for each element of $G$. A group action
    of $G$ on $S$ is then a functor from $BG$ to the category
    $\mathsf{Set}$, whose objects are sets and whose morphisms are
    functions, that sends the single object of $BG$ to the set
    $S$. The fixed point set $S^G$ along with its inclusion in $S$
    then arises as the limit
    (in the category theory sense)
    of this functor.

From here it is easy to see what an ordinary global symmetry of a
    topological field theory should be. Topological field theories do
    not form a set, but rather an $\infty$-groupoid, which we
    generically denote by
    $\gTFT$. To discuss a $G$-symmetry of a topological
    field theory $Z \in \gTFT$ we should therefore first give a
    $G$-action on the $\infty$-groupoid $\gTFT$ of topological field
    theories of interest, \emph{i.e.}\ an $\infty$-functor $BG \to
    \mathsf{Gpd}_\infty$ sending the unique object of $BG$ to $\gTFT$,
    and then the $\infty$-limit $\gTFT^{hG}$ of this functor should
    be thought of as an $\infty$-groupoid of  topological field theories equipped with
    $G$-symmetry. The $\infty$-functor $\gTFT^{hG} \to \gTFT$ sends a topological field theory
    equipped with a $G$-symmetry to the underlying topological field
    theory. It
    need not be either essentially surjective or injective (unlike the inclusion $S^G \hookrightarrow S$), reflecting
    the fact that not every theory need admit a symmetry for the
    given action, and that if it does the symmetry need not be
    unique.

Now the advantage of working exclusively with $\infty$-groupoids, as
we have done, becomes clear: it is that the category of such is
equivalent to the category of spaces, and the categorical
constructions we have just described have down-to-earth
interpretations in terms of homotopy theory of topological spaces and the familiar tools of algebraic topology can
be used to study them. 

The following, then, is a reformulation of the above in terms of topological spaces. Firstly, the $\infty$-groupoid $BG$ corresponds to the classifying space $BG$ of the group $G$, considered as a pointed space (this excuses the clash of notation). Then, giving an action of $G$ on the $\infty$-groupoid $\gTFT$ of topological field theories corrresponds to giving a commutative square of spaces
\begin{equation*}
	\begin{tikzcd}
	\gTFT \arrow{r}{}\arrow{d}{}& 	E \arrow{d}{\pi} \\
	\{*\} \arrow{r}{} & BG
	\end{tikzcd}
\end{equation*}
which is a homotopy pull-back, \emph{i.e.}\ giving a fibration $\pi : E \to
BG$ of spaces, along with an identification $\pi^{-1}(*) = \gTFT$ of
the fibre over the basepoint $* \in BG$ with $\gTFT$.
Finally, the
$\infty$-groupoid $\gTFT^{hG}$ corresponds to the space of sections $s
: BG \to E$ of the fibration $\pi$, and the forgetful morphism
$\gTFT^{hG} \to \gTFT$ corresponds to the map sending a section $s$ to
the point $s(*) \in \gTFT = \pi^{-1}(*)$.

  As we will see in \S \ref{sec:egmext} in the examples with $d=1$, corresponding to
  quantum mechanics, these notions naturally give rise to global
  symmetries with the usual physical properties. In particular, the
  oriented topological field theories in $d=1$ correspond to finite
  dimensional vector spaces, with equivalences given by linear
  isomorphisms. For the trivial action of $G$ on this groupoid,
  the groupoid of homotopy fixed points has objects that are finite-dimensional
  representations of $G$ and morphisms that are invertible
  $G$-equivariant linear maps. For non-trivial actions, we obtain both
  projective and semi-linear representations, and more
    besides. 

  An important subtlety is the following. In defining the notion of a
  global symmetry, we did not take the symmetric monoidal structure of
  topological field theories into account. We could instead have
  defined a
  group action to be an $\infty$-functor from $BG$ to the
  $(\infty,1)$-category of {\em symmetric monoidal}
  $\infty$-groupoids. This makes a difference when we try to take the
  limit, since $\gTFT^{hG}$ must itself then be a symmetric monoidal
  $\infty$-groupoid.
  This would exclude,
  for example, projective representations, the category of which (for a specified
  cocycle) does
  not have the necessary monoidal structure. Since these are
    well-known to occur as global symmetries in Nature, we consider
    our construction to be the appropriate one. 
   %%%%%%%%%%%%%%%%%%%%%%%%%%%%%%%%%%%%%%
    \subsection{Generalized global symmetries\label{sec:gglo}}
    %%%%%%%%%%%%%%%%%%%%%%%%%%%%%%%%%%%%%%
    Our formulation of ordinary global symmetries of topological field
    theories makes it easy to extend to generalized
    symmetries. Indeed, in the categorical language, the group $G$ is
    considered as a 1-truncated $\infty$-groupoid with a single object. So
    the only non-identity morphisms are the 1-morphisms, and these
    correspond to the elements of $G$. 
The only change we need to make to consider generalized symmetries is
to relax the requirement that our $\infty$-groupoid be
$1$-truncated. So it may now have invertible morphisms at all levels,
and these give rise, albeit indirectly, to higher-form
symmetries.

On the homotopy theory side, an $\infty$-groupoid with a single object
corresponds to a pointed connected topological space. Every such space has the homotopy type of
the classifying space $BG$ of some topological group $G$, so we
  continue to refer to it as such. It is
important to note, however, that $G$ is not necessarily unique. For
example, any connected Lie group has a maximal compact subgroup, and
the embedding is both a homomorphism and a homotopy equivalence. Since
the classifying space construction can be made functorial, we conclude
that the classifying space of any connected Lie group has the same
homotopy type of a maximal compact subgroup.\footnote{We will use this fact
later when we discuss generalized gauge symmetries to replace
general linear groups by orthogonal groups.}

With this change made, everything goes through as before. A
generalized group action of $G$ on $\gTFT$ is again an $\infty$-functor $BG \to
\mathrm{Gpd}_\infty$ sending the unique object of $BG$ to $\gTFT$, or equivalently, on the homotopy theory side, a fibration $\pi : E \to BG$ with fibre $\gTFT$ over the basepoint,
and $\gTFT^{hG}$ is the limit of this $\infty$-functor, or
equivalently the space of sections of $\pi : E \to BG$.

Thus we have an extrinsic notion of generalized global
  symmetry. Let us now give, as we did earlier for ordinary symmetries, an intrinsic notion, and connect the two.
{As we saw in the previous section, for any ordinary category, there
  is a natural notion of the symmetry of an object, given by the group of
  automorphisms of that object, with multiplication given by
  composition of morphisms. (For an object in a groupoid, such as a
  TFT, every morphism is an isomorphism, so we can
  equivalently consider the endomorphisms.)
For an object $Z$ in an $\infty$-groupoid $\gTFT$, the morphisms from that
  object to itself themselves form an $\infty$-groupoid. The
  corresponding space is the homotopy pullback of $\{Z\} \hookrightarrow
  \gTFT \hookleftarrow \{Z\}$, \emph{i.e.}\ the loop space
  $\Omega_Z(\gTFT)$. To see that this is sensible, note that an object
  in the
  fundamental $\infty$-groupoid of the space $\gTFT$ is a point $Z \in \gTFT$
  and a 1-morphism from $Z$ to $Z$ is a path from $Z$ to $Z$ in $\gTFT$,
  {\em i.e.} a loop at $Z$. Now, $\Omega_Z(\gTFT)$ does not quite have
  the structure of a group,\footnote{It is, however, a theorem that every loop space
    has the homotopy type of a topological group.} but rather that of an $H$-group. That is,
  it has a multiplication (given by concatenation of loops) and an
  identity (given by the constant loop at $Z$), such that the usual group axioms are obeyed
  up to homotopy.
  
To see the relation between the extrinsic and intrinsic
  symmetries, consider the special case of a trivial fibration $E =
  \gTFT \times BG$. A section of this is simply a map $BG \to
  \gTFT$. It sends $\ast \in BG$ to some $Z \in \gTFT$ and looping we
  get a map $\Omega BG \to \Omega_Z(\gTFT)$ and consequently
  homomorphisms $\pi_q(G) \simeq \pi_{q+1}(BG) \to
  \pi_{q+1}(\gTFT,Z)$, corresponding to actions of $q$-form symmetry
groups $\pi_{q}(G)$ on the theory $Z$.

Passing back to the category theoretic side, we see that $\pi_0(G)$
acts by transformations, $\pi_1(G)$ acts by modifications, and
so on. Now, part of the data of a transformation is a 1-morphism in
the target for each object in the source, a 2-morphism in the target
for each 1-morphism in the source, {\em \&c}, whereas the data of a modification is a
2-morphism in the target for each object in the source, {\em \&c},
{\em \&c}. Spacetime evolutions are associated to $d$-morphisms, and
we see that then the $q$-form symmetries act on what the topological field
theory associates to manifolds of codimension $q+1$, exactly as we
expect for $q$-form symmetries.

Moreover, we see that a non-trivial $q$-form symmetry can only arise
for a
topological field theory that has been extended at least $q+1$ times.
 (Our convention is that a maximally  unextended theory is a functor  out of a 0-category,
{\em i.e.} a
set, whose objects are
closed $d$-manifolds.)
So for a topological field theory formulated using ordinary category
theory ({\em i.e.} once extended, according to the convention just
given), we can have at most an ordinary symmetry. But maximally
extended theories, in which locality is fully manifest, can have
$q$-form symmetries for all $q \leq d-1$.\footnote{As we shall see in
  an example in
  \ref{sec:egunext}, at the level of the action it makes sense to
  speak of $q$-form symmetries acting on $q$-times extended
  theories. But the resulting fixed points represent properties,
  rather than structures, of the corresponding theories.}
%%%%%%%%%%%%%%%%%%%%%%%%%%%%%%%%%%%%%%
\section{Generalized gauge symmetries\label{sec:gau}} 
%%%%%%%%%%%%%%%%%%%%%%%%%%%%%%%%%%%%%%
We now wish to describe generalized gauge symmetries of topological
field theories. An ordinary gauge theory corresponds to
equipping spacetime $M$ with a principal $G$-bundle, or equivalently a
map from $M$ to $BG$. 
Since we saw in the last
Section that a generalized global symmetry can be obtained by
replacing the abstract group $G$ by a topological group, it is natural
to suppose that the same is true in the gauge case.

As we shall see, it is fruitful to do something more general than
equip manifolds with maps to $BG$. Indeed, as well as giving us a
notion of gauge symmetries that act non-trivially on spacetime, it
allows us to subsume the notion of spacetime structures, such as an
orientation or a spin structure. We refer to gauge symmetries that act
trivially on spacetime as internal gauge symmetries. 

This construction closely follows \cite{Lurie:2009keu}, though the
interpretation in terms of generalized symmetries is
presumably new.

Letting $X$ be a topological space, and $\xi$ a rank
$d$ real-vector bundle over $X$, we define the $p$-category
$^{(X,\xi)}\Bord_{d,p}$ as follows: A $(p-k)$-morphism, for $0\le k
\le p-1$ is a triple $(M,f,s)$ consisting of:
   \begin{itemize}
   \item a $(d-k)$-dimensional manifold $M$, with boundary, corners,
     {\em \&c};
   \item a continuous map $f:M\rightarrow X$;
   \item an isomorphism of real vector bundles  $s: TM
       \oplus \underline{\bbR^{k}}
     \rightarrow f^\ast\xi$, where $TM$ is the tangent bundle of $M$, $\underline{\bbR^{k}}$ is the
     trivial rank-$k$ real vector bundle over $M$, and $\oplus$ denotes
     the Whitney sum of bundles.
   \end{itemize}
 For $p$-morphisms, we take the equivalence class of such triples up
   to structure- and corner- preserving diffeomorphisms.

One source of $(X, \xi)$'s is as follows: if $G$ is a topological group and $\chi : G \to O(d)$ is a $d$-dimensional representation, we can take $X = BG$ and $\xi$ to be the vector bundle $EG \times_G \bbR^d$ over $BG$. 

Instead of writing
 $^{(BG,EG\times_G \bbR^d)}\Bord_{d,p}$, we denote this category by
 ${^G\Bord_{d,p}}$, with the homomorphism $\chi$  left implicit.
 Similarly, by ${^G\TFT_{d,p}}$ we denote the
   $\infty$-groupoid of symmetric monodial functors from ${^G\Bord_{d,p}}$ to some
  symmetric monodial $p$-category $\mathsf{C}$. 

Some relevant examples are: {\em (i)} ${^\ast\TFT_{d,p}}$ corresponds
  to framed TFTs (where $\ast$ denotes the group with one
  element); {\em (ii)}  ${^{SO(d)}\TFT_{d,p}}$, with $\chi$ the obvious inclusion, corresponds to oriented
  TFTs, since our conditions correspond to a reduction of structure
  group from $SO(d) \to O(d)$; and {\em (iii)}  ${^{O(d)}\TFT_{d,p}}$,
  with $\chi$ the identity
    map, corresponds to unoriented
    TFTs, which we earlier denoted simply $\TFT_{d,p}$.

 A notion of equivalence of tangential structures is naturally built
 in as follows. A map $\xi \to \xi^\prime$ of vector bundles
 which covers a homotopy equivalence $X \to X^\prime$ and induces a
 linear isomorphism on fibres, gives an equivalence 
 $^{(X,\xi)}\Bord_{d,p} \overset{\sim}\to {^{(X^\prime,\xi^\prime)}\Bord_{d,p}}$ of categories, and therefore an equivalence $^{(X^\prime,\xi^\prime)}\TFT_{d,p} \overset{\sim}\to {^{(X,\xi)}\TFT_{d,p}}$ of spaces of TFTs.
%%%%%%%%%%%%%%%%%%%%%%%%%%%%%%%%%%%%%%
\subsection{Globalization maps\label{sec:gau2glo}} 
%%%%%%%%%%%%%%%%%%%%%%%%%%%%%%%%%%%%%%
We now wish to make a connection between generalized global symmetries
and generalized gauge symmetries and to discuss possible generalized `t~Hooft
anomalies, {\em i.e.} obstructions to gauging generalized global
symmetries.

As we will see, this question of physics has a natural mathematical
formulation in terms of {\em globalization maps} relating spaces of
theories with various combinations of gauged and global symmetries.
These globalization maps formalize, in the topological field theory context, the physicist's notion (for Lie
group gauge symmetries of  theories on spacetime $\bbR^d$) of `turning off the
gauge field'. The issue of `t~Hooft anomalies can then be broken down
into whether a suitable globalization map exists and, if so, whether
it surjects (on $\pi_0$). For maximally-extended theories, the
cobordism hypothesis guarantees the latter.

Let us first try to develop some intuition for globalization
maps by describing the simplest case in which we have an internal
gauge symmetry $G$ ({\em i.e.} the homomorphism
$\chi$ maps $G$ to the identity element in $O(d)$) which we
wish to globalize. We can achieve this by
restricting a topological field theory to spacetime manifolds
equipped with the trivial $G$-bundle, which defines a functor $^G\TFT_{d,p}
\to ^\ast\TFT_{d,p}$, and then consider the effect of bundle
automorphisms, which allows us to factor $^G\TFT_{d,p}
\to ^\ast\TFT_{d,p}$ through  $^\ast\TFT_{d,p}^{hG}$, where the
action of $G$ on $^\ast\TFT_{d,p}$ is the trivial one. (In the
physicist's lingo, we switch off the gauge field and do a constant
gauge transformation.)

More generally, we might want to retain some 
some normal
subgroup of $G$ 
  as a gauge symmetry (or as a spacetime structure), or preserve an existing
  global symmetry, or both. The following construction allows us
  to cover all of these possibilities, and more besides.
  %%%%%%%%%%%%%%%%%%%%%%%%%%%%%%%
  %%%%%%%%%%%%%%%%%%%%%%%%%%%%%%%

We consider the following data: a tangential structure $(X, \xi)$ and
a fibration $\Pi : X \to B$. For each point $b \in B$ we then have a space $X_b := \Pi^{-1}(b)$ with a vector bundle $\xi_b := \xi\vert_{X_b}$ on it, so we think of this data as a continuous family of tangential structures $\{(X_b, \xi_b)\}_{b \in B}$ parameterised by $B$. To this data we may associate the family of bordism $(\infty,p)$-categories $\{^{(X_b, \xi_b)}\Bord_{d,p}\}_{b \in B}$, and by applying the functor $\mathrm{Fun}^\otimes(-, \mathsf{C})$ to each member of this family we obtain a family of $\infty$-groupoids $\{^{(X_b, \xi_b)}\TFT_{d,p}\}_{b \in B}$ parameterised by $B$. Equivalently, we have a space $^{(X, \xi; \Pi)}\TFT_{d,p}$ and a fibration
$$\tau : \, ^{(X, \xi; \Pi)}\TFT_{d,p} \longrightarrow B$$
such that $\tau^{-1}(b) = \, ^{(X_b, \xi_b)}\TFT_{d,p}$.

The inclusions $i_b : X_b \to X$ are by definition covered by bundle isomorphisms $\xi_b \to i_b^* \xi$, so we can canonically consider any manifold equipped with a $(X_b, \xi_b)$-structure as being equipped with a $(X, \xi)$-structure: this defines symmetric monoidal functors $(i_b)_* : \, ^{(X_b, \xi_b)}\Bord_{d,p} \to \, ^{(X, \xi)}\Bord_{d,p}$ and hence restriction functors 
$$i_b^* : \, ^{(X, \xi)}\TFT_{d,p} \longrightarrow \, ^{(X_b, \xi_b)}\TFT_{d,p}.$$
 Thus any topological field theory defined for $(X, \xi)$-manifolds provides a theory for $(X_b, \xi_b)$-manifolds, varying continuously with $b \in B$. That is, there is a map
$$\Pi_* : \, ^{(X, \xi)}\TFT_{d,p} \longrightarrow \{\text{Sections of } \tau : \, ^{(X, \xi; \Pi)}\TFT_{d,p} \to B\}.$$
This map is contravariantly functorial in the data $(\Pi : X \to B, \xi)$. In particular, if $\Gamma$ is a group of symmetries of this data then it acts on source and target of this map, and we can further take the ($\infty$-)fixed points for these $\Gamma$-actions.

The most important source of examples for us will arise from having a group extension
$1  \to K \to G \overset{q}\to Q \to 1$ and a representation $\chi : G \to O(d)$, then taking $X=BG$, $B=BQ$, $\Pi = Bq : BG \to BQ$, and $\xi = EG \times_G \bbR^d$. In this case the construction gives a homotopy pull-back square
\begin{equation*}
	\begin{tikzcd}
	{^{K}\TFT_{d,p}} \arrow{r}{}\arrow{d}{}& 	{^{(G; q)}\TFT_{d,p}} \arrow{d}{\tau} \\
	\{*\} \arrow{r}{} & BQ,
	\end{tikzcd}
\end{equation*}
which as usual corrresponds to an ($\infty$-)$Q$-action on ${^{K}\TFT_{d,p}}$, along with a map
$$\Pi_* : \, ^{G}\TFT_{d,p} \longrightarrow \{\text{Sections of } \tau : \, ^{(G; q)}\TFT_{d,p} \to BQ\},$$
where the latter corresponds to the ($\infty$-)fixed points of the
$Q$-action, and might equally well be denoted by
${^{K}\TFT_{d,p}^{hQ}}$. In physics terms, this corresponds to
  passing from a (generalized, not necessarily internal) gauge symmetry $G$ to a
  normal subgroup $K$, such that the quotient group $Q$ becomes a
  global symmetry.

A notable example comes from the degenerate extension $1 \to 1 \to G \overset{\mathrm{Id}}\to G \to 1$. In this case the homotopy pull-back square
\begin{equation*}
	\begin{tikzcd}
	{^{*}\TFT_{d,p}} \arrow{r}{}\arrow{d}{}& 	{^{(G; \Pi)}\TFT_{d,p}} \arrow{d}{\tau} \\
	\{*\} \arrow{r}{} & BG
	\end{tikzcd}
\end{equation*}
can be identified with that given by the $G$-action via $\chi$ on ${^{*}\TFT_{d,p}}$ by the symmetries of the tangential structure $(*, \bbR^d)$, and so the map in question is
$$\mathrm{Id}_* : \, ^{G}\TFT_{d,p} \longrightarrow \{\text{Sections
  of } \tau : \, ^{(G; \mathrm{Id})}\TFT_{d,p} \to BG\} =
{^{*}\TFT_{d,p}^{hG}}.$$ Physically, we have turned all of the gauge
symmetry into a global symmetry (of $d$-framed TFTs).
%%%%%%%%%%%%%%%%
%%%%%%%%%%%%%%%%%
%%%%%%%%%%%%%%%%%%%%%%%%%%%%%%%%%%%%%%
\subsection{Anomalies and the cobordism hypothesis} 
%%%%%%%%%%%%%%%%%%%%%%%%%%%%%%%%%%%%%%
The globalization map allows us to discuss the notion of `t~Hooft
anomalies ({\em i.e.} global (generalized) symmetries that can't be gauged) in a
precise way. Indeed, we see that to be anomaly-free, a theory $Z \in
\gTFT$ must be in the image of some globalization map. This condition
can be violated in two ways.

Firstly, the space $\gTFT$ in
  which $Z$ lives may not be the target of any globalization map; if
  so we say we have a {\em metaphysical anomaly}, since the anomaly
  afflicts the entire space of TFTs.  Indeed, to be the target of a globalisation
functor,  the action of this global symmetry has to be of a special
kind, namely it must act via symmetries of the tangential
structure. An example of such an anomaly, as we will later see,
is given by representations of topological
field theories in $d=1$ that are genuinely projective, {\em i.e.} those that correspond to cohomologically non-trivial
group cocycles.\footnote{More generally, any space of TFTs with global
symmetry that does not admit a suitable symmetric monoidal structure
must be anomalous.} 

  Secondly, we may have a
  globalization map, but $Z$ may not be in its image. We call this an
  {\em unphysical anomaly}, because it cannot arise in theories that
  are maximally extended, {\em ergo} fully local, as we believe
  theories of physics should be.
This follows by the cobordism hypothesis, whose proof is sketched in
\cite{Lurie:2009keu}, which implies that the map
$$\pi_* : \, ^{(X, \xi)}\TFT_{d,d} \longrightarrow \{\text{Sections of
} \tau : \, ^{(X, \xi; \pi)}\TFT_{d,d} \to B\}$$
is a weak homotopy equivalence.

So (up to equivalence), any theory in the target corresponds to a
unique theory in the source, meaning that neither `t~Hooft anomalies
nor `t~Hooft ambiguities (by which we mean multiple gauge theories with the
same image under globalization) can arise in this way. But for non-maximally extended theories where the cobordism hypothesis
 does not apply, we may also find that the globalization maps fail to
 be either surjective or injective (on $\pi_0$), leading to what we call {\em
   unphysical} 't~Hooft anomalies or ambiguities, respectively.

 It is natural to ask whether one can also have {\em metaphysical
   ambiguities}, in the sense that there exist globalization maps from
 multiple sources to a given target. But to give this concept any
 teeth, one would first need to impose some coarse notion of
 equivalence on tangential structures, presumably based on considerations from physics. If not, then for underlying spaces of
 TFTs that are homotopy $n$-types, one will always find ambiguities
 between tangential structures whose $X$s are equivalent as $n$-types.
    %%%%%%%%%%%%%%%%%%%%%%%%%%%%%%%%%%%%%%
\section{Maximally-unextended theories\label{sec:egunext}} 
%%%%%%%%%%%%%%%%%%%%%%%%%%%%%%%%%%%%%%
We begin our discussion of examples by considering a case which,
although uninteresting as far as physics is concerned, nevertheless
illustrates the mathematics well enough. To wit, we consider
TFTs that are maximally unextended, in that the (truncated) bordism category is
a $0$-category, or set, whose objects are diffeomorphism classes of
closed $d$-manifolds. 

Such theories contain no physics, because they contain
no relations (beyond those implied by the symmetric monoidal
structure) between observables: a theory is specified by its values on diffeomorphism
classes of connected manifolds, and those values are independent of
one another. Nevertheless, the constructions described in previous
sections can be carried out.

On the homotopy theory side, the space $\gTFT$ has the homotopy type
of a discrete space (so each of its connected components is
contractible). We claim that an $\infty$-action of $G$ then corresponds to
the usual notion of a set-theoretic action of $\pi_0(G)$ on $\pi_0(\gTFT)$, and a homotopy fixed point corresponds to a
set-theoretic fixed point. To see this, note that as the notion of $\infty$-action is intrinsically homotopy-invariant, there is no loss of generality in replacing $\gTFT$ with the homotopy equivalent discrete space $\pi_0(\gTFT)$. Then a fibration $\pi : E \to BG$ with fibre $\pi_0(\gTFT)$ is a covering space, so is determined by the monodromy action of $\pi_0(G) = \pi_1(BG, *)$ on $\pi^{-1}(*) = \pi_0(\gTFT)$. A homotopy fixed point of this action is by definition a section $s$ of $\pi$. As $BG$ is path-connected, by the uniqueness of lifts to covering spaces such a section is uniquely determined by $s(*) \in \pi^{-1}(*) = \pi_0(\gTFT)$. Considering $s$ as a map of covering spaces from the trivial covering space $id : BG \to BG$ to $\pi: E \to BG$, we see that $s(*) \in \pi_0(\gTFT)$ must be invariant under the monodromy of $\pi$, \emph{i.e.} \ must be a $G$-fixed point. This identifies $\gTFT^{hG} \simeq \pi_0(\gTFT)^{hG} = \pi_0(\gTFT)^G$, as claimed.
%%%%%%%%%%%%%%%%%%%%%%%%%%%%%%%%%%%%%%%%%%%%%%
\subsection{$d=1$}
%%%%%%%%%%%%%%%%%%%%%%%%%%%%%%%%%%%%%%%%%%%%%%
Things are particularly simple when $d=1$, where $^\ast\TFT_{1,0}$
is equivalent to the space $\Bbbk$, equipped with the discrete
topology.
     To show this, observe that the source ${^\ast\Bord_{1,0}}$
     is the symmetric monoidal $\infty$-groupoid consisting of finite disjoint unions of framed circles. The corresponding homotopy type is the free $E_\infty$-algebra on the space
  $$\{\text{framings of $S^1$}\} /\!\!/ \mathrm{Diff}(S^1).$$
  The space of framings of $S^1$ is the same as the space of orientations of $S^1$, and consists of two contractible path components. The action of an orientation-reversing element of $\mathrm{Diff}(S^1)$ interchanges these components, so the resulting homotopy type is $B\mathrm{Diff}^+(S^1)$, the classifying space of the group of orientation-preserving diffeomorphisms of $S^1$, which in turn is equivalent to $BSO(2) \simeq \mathbb{CP}^\infty$. The target category is obtained by looping
  ${\Vect}$, so it is the set of linear maps $\Bbbk \to
  \Bbbk$ with the symmetric monoidal structure given by tensor product, which is isomorphic to $\Bbbk$ itself with symmetric monoidal structure given by multiplication. It follows that $^\ast
  \TFT_{1,0}$ is the space of continuous maps $B\mathrm{Diff}^+(S^1)
  \to \Bbbk$, which as $B\mathrm{Diff}^+(S^1)$ is connected is simply
  isomorphic to the space $\Bbbk$ with the discrete topology. In other
  words, such a theory is determined by its value on any framed
  circle, and this value can be chosen freely.\footnote{We shall later see that the extended theories in $d=1$ have
values on a framed circle given by the trace of the identity map on
some finite-dimensional vector space, so
the extendable theories are those which take the
value $\sum_{i=1}^n 1 \in \Bbbk$, for some $n \in \{0,1,2,\dots\}$.}
%%%%%%%%%%%%%%%%%%%%%%%%%%%%%%%%%%%%%%%%%%%%%%

Now let us consider the possible gauge symmetries. 
For simplicity,
  we consider here only internal gauge symmetries, so we take a topological group $G$ and the trivial representation $\chi : G \to O(1)$, and describe  $^G \TFT_{1,0}$. Now the source is the symmetric monoidal $\infty$-groupoid ${^G\Bord_{1,0}}$, whose corrresponding homotopy type is the free $E_\infty$-algebra on the space
    $$\{\text{framings of $S^1$, $f : S^1 \to BG$}\} /\!\!/ \mathrm{Diff}(S^1).$$
Following the discussion above we only need to understand the set of
path-components of this space, which is the same as $\{\text{$f : S^1
  \to BG$}\} /\!\!/ \mathrm{Diff}^+(S^1)$, and as the group
$\mathrm{Diff}^+(S^1)$ is connected the path components of this are
identified with $\pi_0 \{\text{$f : S^1 \to BG$}\}$, or in other words
with the set $\mathrm{Conj}(\pi_0 G)$ of conjugacy classes of elements
of $\pi_0 G$. Thus $^G \TFT_{1,0}$ is identified with the set of
$\Bbbk$-valued functions on this set, \emph{i.e.}\ the $\Bbbk$-valued class
functions on $\pi_0 G$.\footnote{The extendable theories are given
    by the class functions that are characters of representations of $\pi_0 G$.}  Thus we might as well take $G$ to be
  discrete in what follows.
%%%%%%%%%%%%%%%%%%%%%%%%%%%%%%%%%%%%%%%%%%%%%%%%
By the general arguments already given, the possible
global symmetries of $^K\TFT_{1,0}$
correspond to fixed points of some action of some discrete group $Q$ on 
the set of class functions $K \to
\Bbbk$, so let us now
consider the globalization maps. Starting from an internal gauge
symmetry based on discrete group $G$ as above, 
every fibration $\Pi$ of tangential structures is equivalent
to a short
exact sequence $\ast \to K \to G \to Q \to \ast$ of groups.
For such a
sequence, there is an action of $Q \simeq G/K$ on the set $C_K$ of
conjugacy classes of $K$ given by $\rho: G/K \times C_K \to C_K: ([g],[k])
\mapsto [gkg^{-1}]$. The fibration $\tau$
corresponds to the $Q$ action on $\Map (C_K,\Bbbk)$ that is induced by $\rho$,
and $\Pi_*$ corresponds to the map $\Map (C_G,\Bbbk) \to \Map (C_K,\Bbbk)^Q$ given by restriction. Since a normal subgroup
of $G$ is a union of conjugacy classes, it is easy to see
that the map $\Pi_*$ always surjects, but injects iff $K=G$. So there
are no unphysical anomalies, but plenty of unphysical ambiguities. 

In contrast, we certainly have metaphysical anomalies, for any $Q$-action
on $\Map (C_K,\Bbbk)$
that is not of the form above ({\em e.g. }one that does not fix the image of
all maps) cannot be gauged.

%%%%%%%%%%%%%%%%%%%%%%%%%%%%%%%%%%%%%%
\section{Maximally-extended theories\label{sec:egmext}} 
%%%%%%%%%%%%%%%%%%%%%%%%%%%%%%%%%%%%%%
We now wish to focus on the case of maximally-extended theories, with
$p=d$. Not only are these the ones of interest to physics (being fully
local), but they also lead to a number of simplifications thanks to
the cobordism hypothesis. Theories that are not maximally extended are
less pleasant and will be studied in \S\ref{sec:egnmext}.

At least in low dimension $d$, maximally-extended theories are fairly
simple to classify and study. In the following subsections, we focus
on the cases of $d=1$, and $d=2$, where we investigate symmetries and
anomalies in more detail. Unfortunately, for $d>2$, there is even less
consensus on what a suitable target category for TFTs might be,
let alone an description of the corresponding space of TFTs.
%%%%%%%%%%%%%%%%%%%%%%%%%%%%%%%%%%%%%%
\subsection{$d=1$\label{sec:egmext1}} 
 %%%%%%%%%%%%%%%%%%%%%%%%%%%%%%%%%%%%%%
Here we will classify group actions on framed (equivalently
     oriented), fully-extended TFTs for $d=1$ valued in $\Vect_\Bbbk$. It is easily
     shown that $^\ast\TFT_{1,1}$ is equivalent to
     the groupoid (\emph{i.e.}\ 1-truncated $\infty$-groupoid) of finite-dimensional vector spaces (including the
     zero-dimensional space) and linear
   isomorphisms.

   Though this case of `topological quantum mechanics' may seem
     rather boring from the dynamical perspective, we will see that it
     admits a rich variety of global symmetries.

 As a topological space, we may take the 1-type
\begin{align}
{^\ast \TFT_{1,1}}=\coprod_{n\geq 0} BGL(n,\Bbbk),
\end{align}
the disjoint union of the classifying spaces of the groups
$GL(n,\Bbbk)$ with the discrete topology. As we have discussed, an ($\infty$-)action of $G$ on $^\ast\TFT_{1,1}$ corresponds to the data of a homotopy pull-back square
\begin{equation*}
	\begin{tikzcd}
	{^{*}\TFT_{1,1}} \arrow{r}{}\arrow{d}{}& 	E \arrow{d}{\pi} \\
	\{*\} \arrow{r}{} & BG.
	\end{tikzcd}
\end{equation*}
As two $GL(n,\Bbbk)$ for different $n$ cannot be
isomorphic~\cite{1971221}, we must have $E \simeq \coprod_{n \geq 0}
E_n$, and assuming $E_n \to BG$ is splittable (if not, the
  space of homotopy fixed points will be empty) we must have $E_n \simeq B\tilde{G}_n$  for topological groups $\tilde{G}_n$ fitting into splittable extensions
        \begin{gather} \label{eq:GLnSES}
	0\to GL(n,\Bbbk)\xrightarrow{\beta} \tilde
        G_n\xrightarrow{\alpha} G \to 0.
      \end{gather}
Having fixed a $G$-action on ${^{*}\TFT_{1,1}}$ as above, we find that
$${^{*}\TFT_{1,1}^{hG}} \simeq \coprod_{n \geq 0} \{\text{splittings $s$ of $\alpha : \tilde{G}_n \to G$}\} /\!\!/ GL(n,\Bbbk),$$
where $/\!\!/$ denotes the action groupoid (or homotopy quotient) for
the $GL(n,\Bbbk)$-action on the set of splittings by
conjugation. As $GL(n,\Bbbk)$ is discrete, splitting such an extension is the same as splitting $\pi_0 \alpha : \pi_0 \tilde{G}_n \to \pi_0 G$, so we may as well suppose that $G$ is discrete.

Group extensions of $G$ by $GL(n,\Bbbk)$ are classified by the
non-abelian group cohomology $H^2(G,GL(n,\Bbbk))$ (see
\emph{e.g.}~\cite{Muller:2018doa,Muller:2020phm}), in which a 2-cocycle is
a pair $\left(\sigma:G\rightarrow
\mathrm{Aut}(GL(n,\Bbbk)),\epsilon:G\times G \rightarrow
GL(n,\Bbbk)\right)$ of functions
satisfying certain conditions.\footnote{To wit (letting $1$ denote the
  identities in both $G$ and $GL(n,\Bbbk)$): $\sigma(1)=\mathrm{id}_{GL(n,\Bbbk)}$, $\epsilon(1,1)=1$, $\sigma(g_1g_2)(N)=\epsilon(g_1,g_2)^{-1} \sigma(g_1)(\sigma(g_2)(N)) \epsilon(g_1,g_2)$, and $\epsilon(g_1,g_2)\epsilon(g_1g_2,g_3)=\sigma(g_1)(\epsilon(g_2,g_3)) \epsilon(g_1,g_2g_3)$.}
 Two pairs $(\sigma,\epsilon)$ and
 $(\sigma^\prime,\epsilon^\prime)$ represent the same cohomology class
 (which we denote $\llbracket \sigma,\epsilon\rrbracket \in H^2(G,GL(n,\Bbbk))$)
 iff there exists a function $t:G\rightarrow GL(n,\Bbbk)$ such that 
 \begin{align}
 	\sigma^\prime(g)(N)&=t(g)\sigma(g)(N)t(g)^{-1},\nonumber\\
 	\epsilon^\prime(g_1,g_2)&=t(g_1)\sigma(g_1)(t(g_2))\epsilon(g_1,g_2)  t(g_1g_2)^{-1},
 \end{align}
 for all $N\in GL(n,\Bbbk)$ and $g,g_1,g_2\in G$.

As described above, we are only interested in short exact sequences which admit a splitting, since
otherwise there will be no homotopy fixed points. 
In terms of non-abelian group
cohomology, a sequence splits iff its cohomology
class has a representative of the form 
$\llbracket \sigma,\mathbb{I}\rrbracket$ where $\mathbb{I}(g_1,g_2)=1$
is the constant map, in which case $\sigma$ is a
homomorphism. 
Indeed given a section $s$ of the sequence, we can define
$\sigma_s:G\rightarrow \mathrm{Aut}(GL(n,\Bbbk)):g\mapsto(N\mapsto
\beta^{-1}(s(g)\beta(N)s(g)^{-1}))$ and $\llbracket \sigma_s,\mathbb{I}\rrbracket$ is the cohomology class of the SES.
This reproduces the well-known fact that a sequence splits iff it is
equivalent to a semi-direct product.

The set of sections of a sequence with representative 2-cocycle
  $(\sigma,\epsilon)$ can be conveniently described as follows: it is
  in bijection with functions $r:G\rightarrow GL(n,\Bbbk)$ that are twisted
  versions of representations, in the sense that
\begin{align}\label{eq:TwistedRep}
	r(g_1)\sigma(g_1)(r(g_2))\epsilon(g_1,g_2)=r(g_1g_2).
\end{align}
From the homotopy quotient, we see that two twisted representations
$r$ and $r^\prime$ are to be considered equivalent if there exists an $M\in GL(n,\Bbbk)$  such that 
\begin{align}\label{eq:Interwine}
	Mr(g)\epsilon(1,g)=r^\prime(g)\sigma(g)(M)\epsilon(g,1)
\end{align}
for all $g\in G$.

In the special case when $\sigma$ and $\epsilon$ are both trivial, Eq.~\ref{eq:TwistedRep} corresponds to the condition for
standard representations, and  Eq.~\ref{eq:Interwine}  the usual
equivalence of representations. More generally, if just $\sigma$  is
the trivial map, then Eq.~\ref{eq:TwistedRep} corresponds to the
condition for projective representations with twisting $\epsilon$,
and, since  $\epsilon(1,g)=\epsilon(g,1)$ in this case,
Eq.~\ref{eq:Interwine}  corresponds to the usual `linear equivalence' of projective representations.

It follows from the above discussion that all we need to describe
  a splittable short exact sequence is a homomorphism
  $\sigma:G\rightarrow \mathrm{Aut}(GL(n,\Bbbk))$. Thus we need to
  describe all possible automorphisms of $GL(n,\Bbbk)$.
  
 These are indeed known,
though complicated \cite{Dieudonne1963}.
In a nutshell, all automorphisms arise from: {\em (i)}
inner automorphisms; {\em (ii)} field automorphisms of $\Bbbk$; {\em
  (iii)} the involution given by taking the inverse transpose; and
{\em (iv)} homomorphisms $\chi: GL(n,\Bbbk) \to
\Bbbk^\ast$.

Let us now spell out some examples of physical interest,
  corresponding to the different types of automorphism of
  $GL(n,\Bbbk)$ described above. We will see that Nature makes use of all but the last
one.

  The usual representations of physics arise from the trivial action,
  which exists for every $G$. Here $\sigma_s$ sends all of $G$ to the
  identity automorphism and we may take $\tilde G_n=GL(n,\Bbbk)
  \times G$ for all $n$. For a given $n$, the space of sections are then in 1--1
  correspondence with homomorphisms $r:G\rightarrow  GL(n,\Bbbk)
  $, {\em i.e.} representations of dimension $n$. A morphism between two homotopy fixed points with representations
  $r_{1}, r_2:G\rightarrow  GL(n,\Bbbk)$ corresponds to an $M\in
  GL(n,\Bbbk)$ such that $M r_1(g)=r_2(g) M$, for all $g\in
  G$, {\em i.e.} the usual notion of equivalence of
  representations.

 More generally, inner automorphisms 
    of $GL(n,\Bbbk)$ give rise to projective
    representations, as described above.
  We remark that the possible occurrence of projective representations in
  physics is usually derived from the axiom that physical states in
  quantum mechanics correspond to rays in Hilbert space. Here we have
  no such axiom (the notion of a ray in a vector space certainly makes sense,
  but it is not clear what `physical' should mean, given that we have
  no way of extracting real numbers that could be interpreted as
  predictions for physics measurements),\footnote{Even for $\Bbbk =
    \bbC$, we have no way to identify $\bbR \subset \bbC$.} but it is
  nevertheless reassuring to see that projective
  representations are nevertheless allowed by the primitive requirements of
  locality and  entanglement that the axioms of topological field
  theories encode.

  When the homomorphism of $G \to \Aut(GL(n,\Bbbk))$ is
    induced by a homomorphism $G \to \Aut(\Bbbk)$, we get semi-linear
    representations.
    An important case for physics occurs when $\Bbbk = \bbC$ and we choose
some involution of $\bbC$ (which defines a complex conjugation in
$\bbC$ relative to the real line, defined as the fixed
point subset). Then we get the antilinear representations, of which
time reversal symmetry is an example. Here, though, we have no notion
of time and no notion of unitarity.

Next consider the inverse transpose automorphism. This has a special
r\^{o}le to play, because it corresponds to the $O(1)$-action on
$^\ast \TFT_{1,1}$. By the cobordism hypothesis, its groupoid of homotopy fixed points should be equivalent to the groupoid of topological field
theories on unoriented manifolds. To see this, let $G=O(1) = \{+1,-1\}$, and let
$\tilde G_n=GL(n,\Bbbk)\rtimes O(1)$, where the semi-direct product is
defined via the multiplication rule
\begin{align}
	(M_2, 1)\cdot (M_1,\pm 1)=(M_2M_1,\pm 1),\quad 	(M_2, -1)\cdot (M_1,\pm 1)=(M_2(M_1^{-1})^T,\mp 1).
\end{align}
A splitting $s$ of the corresponding extension is specified by its value $A \in
GL(n,\Bbbk)$ on $-1 \in O(1)$. By considering $s(-1 \cdot -1) = s(1)$, we find
that $A^T=A$. So splittings correspond to non-degenerate symmetric bilinear
forms. 
A morphism between the splittings corresponding to $A_{1}, A_2\in
GL(n,\Bbbk)$ is
given by an $M\in GL(n,\Bbbk)$ such that $MA_1M^T=A_2$, which
corresponds to the usual notion of equivalence of non-degenerate
symmetric bilinear forms
(\emph{i.e.} $^{*}\TFT_{1,1}^{hO(1)}$ is equivalent to the groupoid of 
finite-dimensional vector spaces equipped with non-degenerate
symmetric bilinear forms and whose morphisms are linear isomorphisms
which preserve the forms under pullback, which is indeed the same as $^{O(1)}\TFT_{1,1}$).

It is difficult to say much more for generic fields $\Bbbk$. However,
for algebraically closed fields (such as $\bbC$) all $A$ are
equivalent to the identity matrix, and their automorphisms are
isomorphic to $O(n,\Bbbk)$. Thus, in this case we have that 
\begin{align}
	^{O(1)}\TFT_{1,1}=\coprod_{n=0}^\infty BO(n,\Bbbk).
\end{align}
When $\Bbbk=\bbR$, Sylvester's law of inertia tells us that, up to equivalence,
the $A$ are given by diagonal matrices whose diagonal entries are $\pm
1$, with automorphisms given by $O(p,q,\bbR)$. Thus we have
\begin{align}
	^{O(1)}\TFT_{1,1}=\coprod_{p=0}^\infty \coprod_{q=0}^\infty BO(p,q, \bbR).
\end{align}

In closing, it is perhaps of interest to speculate whether there might exist
yet more ways of realizing symmetries, as yet unknown to physics. At
least when $\Bbbk = \bbC$, this seems unlikely. All the field
automorphisms of $\bbC$ are either of order two, so define a real line
and a complex conjugation as needed to define the values of
physical observables, or are of infinite order. But complex
conjugation is the  
only automorphism of $\bbC$ considered as an
$\bbR$-algebra. The only non-trivial
automorphisms obtained from homomorphisms $\chi: GL(n,\bbC) \to
\bbC^\ast$ correspond to powers of the determinant map. For a
non-zero power, the
resulting automorphism of $GL(n,\bbC)$ generates a subgroup isomorphic
to $\bbZ$, so doesn't admit a non-trivial automorphism from a finite
group.
%%%%%%%%%%%%%%%%%%%%%%%%%%%%%%%%%%%%%%
\subsection{$d=2$\label{sec:eg2}}
%%%%%%%%%%%%%%%%%%%%%%%%%%%%%%%%%%%%%%
We now carry out an analysis of the possible generalized global
symmetries of framed TFTs in $d=2$ valued in $\Alg_\Bbbk$. 
As we will see, this can be done in full, at least when the field $\Bbbk$
is separably closed, though the result is somewhat
complicated.

Keeping $\Bbbk$ general to begin with, the cobordism hypothesis \cite{Lurie:2009keu} states
that the bicategory of framed TFTs valued in $\Alg_\Bbbk$ is given by
the core of the fully-dualizable objects in $\Alg_\Bbbk$, where 
a $\Bbbk$-algebra $A$ is fully-dualizable iff
          it is separable, meaning that $A
           \otimes_{\Bbbk} \mathbb{K}$ is finite dimensional and
           semisimple for every field extension $\mathbb{K} \supset
           \Bbbk$.
Choosing $\mathbb{K} =
           \Bbbk$, we see that $A$ itself is finite dimensional and
           semisimple, so we can apply the Artin--Wedderburn
           theorem. To do so, we
           need to know the finite-dimensional division algebras over $\Bbbk$. Since $A$
           is separable, these must be too, so we are looking for
           finite-dimensional division algebras whose centres are
           finite-dimensional separable field extensions of the field
           $\Bbbk$.

           In general, this is difficult, as can be seen by
           considering the extreme case in which $\Bbbk$ is perfect (meaning that every algebraic extension is
  separable), in which case every finite-dimensional division algebra
  over $\Bbbk$ is valid and being separable is equivalent to being
  finite dimensional and semisimple. This case includes all fields of characteristic zero and all finite
  fields, so probably every field that could be of interest to
  physicists. But finding the division algebras, even for a specific $\Bbbk$, is a hard (though
  well-studied) problem. 

Instead we choose to focus here on the opposite extreme in which 
$\Bbbk$
  is separably closed (meaning that no algebraic
  extension is separable),
  in which case the only division algebra is $\Bbbk$ itself. This case
  includes the one of most interest, namely $\Bbbk
  =\bbC$ (which is algebraically closed so separably closed). 
\subsubsection*{Separably closed fields}
%%%%%%%%%%%%%%%%%%%%%%%%%%%%%%%%%%%%%%
For separably closed fields $\Bbbk$, the Artin--Wedderburn theorem tells us
  that every separable algebra is isomorphic as an algebra to a finite
  product of matrix algebras over $\Bbbk$, but every such algebra is
  Morita equivalent to $\Bbbk^n$ for some positive integer $n$.
  Generalizing the arguments
    in \cite{davidovich_2011}, one
    finds
\begin{align}\label{eq:DavidovichEq}
{^\ast\TFT_{2,2}} \cong \coprod_{n \geq 1} E S_n \times_{S_n} K(\Bbbk^\ast, 2)^{\times n}
\end{align}
where the permutation group $S_n$ acts on $K(\Bbbk^\ast,
  2)^{\times n}$ by permuting the factors. (We remark that unlike for
  theories in $d=1$, the sum here starts from $n=1$, since there is no
  zero-dimensional algebra.)

As usual, an action of $G$ on ${^\ast\TFT_{2,2}}$ is described by a homotopy pull-back square
\begin{equation*}
	\begin{tikzcd}
	{^\ast\TFT_{2,2}} \arrow{r}{}\arrow{d}{}& 	E \arrow{d}{\pi} \\
	\{*\} \arrow{r}{} & BG.
	\end{tikzcd}
\end{equation*}
As the path-components of ${^\ast\TFT_{2,2}}$ have non-isomorphic fundamental groups (namely the distinct symmetric groups), the $G$-action preserves path-components and so there is a corresponding decomposition $E = \coprod_{n \geq 1} E_n$. As we are only interested in $G$-actions which admit homotopy fixed points, writing $X_n = E S_n \times_{S_n} K(\Bbbk^\ast, 2)^{\times n}$  we are therefore looking for homotopy fibre sequences
\begin{equation}\label{eq:XnExtension}
X_n \longrightarrow E_n  \overset{\pi}\longrightarrow BG
\end{equation}
which admit a section. Using that $\pi_1(X_n) = S_n$ and $\pi_2(X_n) = (\Bbbk^\ast)^{\oplus n}$ with $S_n$-module structure given by permuting the summands, we can understand such a homotopy fibre sequence by developing the diagram
\begin{equation}\label{eq:BigDiagram}
	\begin{tikzcd}
	K((\Bbbk^\ast)^{\oplus n}, 2) \arrow{d}{} \arrow[r, equals]& K((\Bbbk^\ast)^{\oplus n}, 2) \arrow{r}{}\arrow{d}{}& 	* \arrow{d}{} \\
	X_n \arrow{r}{} \arrow{d}{}& E_n \arrow{r}{\pi} \arrow{d}{p} & BG \arrow[d, equals] \arrow[bend left=30, swap]{l}{s}\\
	BS_n \arrow{r}{}& E_n' \arrow{r}{\pi'} & BG \arrow[bend left=30, swap]{l}{s_0 = ps}
	\end{tikzcd}
\end{equation}
in which all rows and columns are homotopy fibre sequences, by letting $\pi'$ be the fibrewise 1-truncation of $\pi$. The bottom row, with a choice $s_0$ of section, is classified by the data of:
\begin{enumerate}[(i)]
\item a homomorphism $\pi_0 G \to \mathrm{Aut}(S_n)$.

\end{enumerate}
Given such a choice, which in particular identifies $E_n' \simeq B(S_n \rtimes G)$, the middle row is classified by the data of:
\begin{enumerate}[(i)]
\setcounter{enumi}{1}

\item a $S_n \rtimes \pi_0 G$-module structure on $(\Bbbk^\ast)^{\oplus n}$ extending the $S_n$-module structure,

\item a class $k \in H^3(B(S_n \rtimes G) ; (\Bbbk^\ast)^{\oplus n})$ which vanishes when restricted to $H^3(BS_n ; (\Bbbk^\ast)^{\oplus n})$.
\end{enumerate}
In order for the resulting $\pi$ to admit a section, this should satisfy
\begin{enumerate}[(i')]
\setcounter{enumi}{2}

\item $k \in H^3(B(S_n \rtimes G) ; (\Bbbk^\ast)^{\oplus n})$ vanishes when restricted to $H^3(BG ; (\Bbbk^\ast)^{\oplus n})$.
\end{enumerate}
In this case the homotopy classes of sections $s$ lifting the given $s_0$ are given by the reasons this class vanishes, \emph{i.e.}\ are a torsor for $H^2(BG ; (\Bbbk^\ast)^{\oplus n})$.

The data in (i) and (ii) can be packaged together as follows. There is a group $\mathrm{Aut}(S_n, (\Bbbk^\ast)^{\oplus n})$ consisting of pairs of a group isomorphism $f_0 : S_n \to S_n$ and a $f_0$-linear module isomorphism $f_1 : (\Bbbk^\ast)^{\oplus n} \to (\Bbbk^\ast)^{\oplus n}$, and (i) and (ii) combined correspond to a homomorphism
$$\phi : \pi_0 G \to \mathrm{Aut}(S_n, (\Bbbk^\ast)^{\oplus n}).$$
To analyse this group, first note that for $n \neq 6$ all automorphisms $f_0$ of $S_n$ are inner, and it is clear that these admit a canonical corresponding $f_1$. On the other hand, for $n=6$ the outer automorphism $f_0$ of $S_6$ does \emph{not} admit a corresponding $f_1$ 
(unless $\Bbbk^\ast$ is trivial), so for the data (ii) to exist, in (i) we must choose a homomorphism $\pi_0 G \to \mathrm{Inn}(S_n)$. This discussion gives us a split extension
$$1 \to \mathrm{Aut}_{\bbZ[S_n]}((\Bbbk^\ast)^{\oplus n}) \to \mathrm{Aut}(S_n, (\Bbbk^\ast)^{\oplus n}) \to \mathrm{Inn}(S_n) \to 1.$$
The kernel in this extension can be interpreted as the subgroup of $GL(n, \mathrm{End}_\bbZ(\Bbbk^\ast))$ consisting of those matrices which centralise the permutation matrices. It is easy to see that these are the invertible matrices which have a common entry at all diagonal positions and another common entry at all off-diagonal positions, \emph{i.e.}\ invertible matrices of the form
\begin{equation}\label{eq:matrix}
\begin{pmatrix}
a+b & b & \ldots & b\\
b & a+b & \ldots & b\\
\vdots & \vdots & \ddots & \vdots\\
b & b & \ldots & a+b\\
\end{pmatrix} \quad a, b \in \mathrm{End}_\bbZ(\Bbbk^\ast)
\end{equation}
Such a matrix is invertible for $n>1$ if and only if both $a$ and $a+n \cdot b$ are invertible in $\mathrm{End}_\bbZ(\Bbbk^\ast)$.

The scope of the data in (iii) (satisfying (iii')) can be analysed by considering the Serre spectral sequence for the bottom row of \eqref{eq:BigDiagram} with $(\Bbbk^\ast)^{\oplus n}$-coefficients. This describes the group $K$ of all possible $k$'s in terms of an exact sequence
\begin{equation}\label{eq:snake}
\begin{tikzcd}
  H^0(BG ; H^2(S_n ; (\Bbbk^\ast)^{\oplus n})) \rar{d_2} & H^2(BG ; H^1(S_n ; (\Bbbk^\ast)^{\oplus n})) \rar
             \ar[draw=none]{d}[name=X, anchor=center]{}
    & K \ar[rounded corners,
            to path={ -- ([xshift=2ex]\tikztostart.east)
                      |- (X.center) \tikztonodes
                      -| ([xshift=-2ex]\tikztotarget.west)
                      -- (\tikztotarget)}]{dll}[at end]{} \\      
  H^1(BG ; H^2(S_n ; (\Bbbk^\ast)^{\oplus n})) \rar{d_2} & H^3(BG ; H^1(S_n ; (\Bbbk^\ast)^{\oplus n}))
\end{tikzcd}
\end{equation}
As a $S_n$-module we have $(\Bbbk^\ast)^{\oplus n} = \mathrm{coInd}_{S_{n-1}}^{S_n} \Bbbk^\ast$, so by Shapiro's lemma we have $H^i(S_n ; (\Bbbk^\ast)^{\oplus n}) \cong H^i(S_{n-1} ; \Bbbk^\ast)$, which may be determined using the Universal Coefficient Theorem and the known low-degree homology of symmetric groups: the result is shown in Table \ref{table:1}. 
\begin{table}[]
\begin{tabular}{l|lllll}
$n$ & 1 & 2 & 3 & 4 & $\geq 5$ \\ \hline
$H^2(S_n ; (\Bbbk^\ast)^{\oplus n})$ & 0 & 0 & $\Bbbk^\ast/(\Bbbk^\ast)^2$ & $\Bbbk^\ast/(\Bbbk^\ast)^2$ & $\Bbbk^\ast/(\Bbbk^\ast)^2 \oplus \Bbbk^\ast[2]$  \\
$H^1(S_n ; (\Bbbk^\ast)^{\oplus n})$ & 0 & 0 & $\Bbbk^\ast[2]$ & $\Bbbk^\ast[2]$ & $\Bbbk^\ast[2]$
\end{tabular}
\caption{$\Bbbk^\ast[2]$ and $\Bbbk^\ast/(\Bbbk^\ast)^2$ denote the kernel and cokernel respectively of the squaring map $(-)^2 : \Bbbk^\ast \to \Bbbk^\ast$.}\label{table:1}
\end{table}
 The group $\mathrm{Aut}(S_n, (\Bbbk^\ast)^{\oplus n})$ acts on $H^i(S_n ; (\Bbbk^\ast)^{\oplus n})$ by functoriality of group cohomology in the group and in the module. The subgroup $\mathrm{Inn}(S_n) \leq \mathrm{Aut}(S_n, (\Bbbk^\ast)^{\oplus n})$ given by the splitting acts trivially on $H^i(S_n ; (\Bbbk^\ast)^{\oplus n})$, because inner automorphisms act trivially on cohomology \cite[Proposition III.8.3]{Brown}. A matrix of the form \eqref{eq:matrix} in the subgroup $\mathrm{Aut}_{\bbZ[S_n]}((\Bbbk^\ast)^{\oplus n}) \leq \mathrm{Aut}(S_n, (\Bbbk^\ast)^{\oplus n})$ acts as $a+n\cdot b \in \End_\bbZ(\Bbbk^\ast)$ on $H^i(S_{n-1} ; \Bbbk^\ast)$, so by the induced map on the $\Bbbk^\ast[2]$ and $\Bbbk^\ast/(\Bbbk^\ast)^2$ in Table \ref{table:1}. This describes the $\pi_0 G$-action on $H^i(S_n ; (\Bbbk^\ast)^{\oplus n})$ with which the cohomology groups in \eqref{eq:snake} are taken.

If $\Bbbk$ is a separably closed field of characteristic $\neq 2$ then it is closed under the formation of square roots, and so $\Bbbk^\ast/(\Bbbk^\ast)^2$ is trivial, and $\Bbbk^\ast[2] = \{\pm 1\}$, which has no automorphisms. Thus for $n \geq 5$ the group $K$ fits into an exact sequence
$$H^0(BG ; \{\pm 1\}) \overset{d_2}\to H^2(BG ; \{\pm 1\})\to K \to H^1(BG ; \{\pm 1\})\overset{d_2}\to H^3(BG ; \{\pm 1\}).$$
(If the $\pi_0 G$-action on $S_n$ is trivial then the $d_2$-differentials are zero, so $K$ is determined up to an extension problem.) If instead $\Bbbk$ has characteristic 2 then $\Bbbk^\ast[2]$ is trivial, so for $n \geq 3$ there is an isomorphism
$$K \overset{\sim}\to H^1(BG ; \Bbbk^\ast/(\Bbbk^\ast)^2),$$
where $\pi_0 G$ acts on $\Bbbk^\ast/(\Bbbk^\ast)^2$ as described in the previous paragraph.

Let us now discuss the space of sections of \eqref{eq:XnExtension}, \emph{i.e.}\ the homotopy $G$-fixed points of the $G$-actions on $X_n$ that we have just described. As we have classified such fibrations with a choice of section $s_0 : BG \to E_n'$ of $\pi' : E_n' = B(S_n \rtimes G) \to BG$, we may as well focus on the space $\Gamma(\pi)_{s_0}$ of sections of $\pi : E_n \to BG$ such that $ps$ is in the path component of $s_0$. Composing with $p$ gives a fibration
$$p_* : \Gamma(\pi)_{s_0} \to \Gamma(\pi')_{s_0}$$
to the space of sections of $\pi'$ in the path component of $s_0$. Precisely as in the previous section, the space $\Gamma(\pi')$ is homotopy equivalent to $\{\text{splittings of $S_n \rtimes \pi_0G \to \pi_0 G$}\}/\!\!/ S_n$. Thus the path-component of $s_0$ is a classifying space for the stabiliser $\mathrm{St}_{S_n}(s_0) \leq S_n$, which may be seen to be the subgroup of elements which centralise $\mathrm{Im}(\pi_0 G \to \mathrm{Inn}(S_n))$. The fibre $F$ of $p_*$ over $s_0$ is the space of sections $s : BG \to E_n$ such that $ps=s_0$. This may be viewed as the space of trivialisations of the (trivial) class $(s_0)^*(k) \in H^3(BG ; (\Bbbk^\ast)^{\oplus n})$, so its set of path components is a torsor for $H^2(BG ; (\Bbbk^\ast)^{\oplus n})$, and
$$\pi_i(F, s) \cong H^{2-i}(BG ; (\Bbbk^\ast)^{\oplus n}) \text{ for } i > 0.$$
The long exact sequence on homotopy groups for the fibration $p_*$ then gives
\begin{align*}
0 \to H^{1}(BG ; (\Bbbk^\ast)^{\oplus n}) &\to \pi_1(\Gamma(\pi), s) \to \mathrm{St}_{S_n}(s_0) \overset{\partial}\to H^{2}(BG ; (\Bbbk^\ast)^{\oplus n}) \to \pi_0(\Gamma(\pi)_{s_0}) \to *\\
H^{0}(BG ; (\Bbbk^\ast)^{\oplus n}) &\overset{\sim}\to \pi_2(\Gamma(\pi), s)
\end{align*}
and all higher homotopy groups of $\Gamma(\pi)$ are trivial. The map $\partial$ is a crossed homomorphism, \emph{i.e.}\ an element of $H^1(\mathrm{St}_{S_n}(s_0) ; H^{2}(BG ; (\Bbbk^\ast)^{\oplus n}))$, and corresponds to restricting $k \in H^3(B(S_n \rtimes G) ; (\Bbbk^\ast)^{\oplus n})$ to the subgroup $\mathrm{St}_{S_n}(s_0) \times G \leq S_n \rtimes G$ and then applying the map
$$\mathrm{Ker}(H^3(B(\mathrm{St}_{S_n}(s_0) \times G); (\Bbbk^\ast)^{\oplus n}) \to H^3(BG; (\Bbbk^\ast)^{\oplus n})) \to H^1(\mathrm{St}_{S_n}(s_0) ; H^{2}(BG ; (\Bbbk^\ast)^{\oplus n}))$$
coming from the Serre spectral sequence for $\mathrm{St}_{S_n}(s_0) \times G \to \mathrm{St}_{S_n}(s_0)$.

As an example, let us consider the trivial $G$-action on $^* \TFT_{2,2}$. A $G$-homotopy fixed point whose underlying topological field theory is an algebra Morita equivalent to $\Bbbk^n$ corresponds to a section $s$ of the trivial fibration $BG \times X_n \to BG$, or in other words to a map $f : BG \to X_n$. In terms of our classification this corresponds to the homomorphism $ \pi_0 G \overset{\pi_1 f}\to S_n \to \mathrm{Inn}(S_n)$ and the $S_n \rtimes \pi_0 G$-module structure on $(\Bbbk^\ast)^{\oplus n}$ induced by $(\sigma, g) \mapsto \sigma \cdot \pi_1 f(g) : S_n \rtimes \pi_0 G \to S_n$ and the usual $S_n$-module structure. As the underlying fibration is trivial, $k=0$ and so the crossed homomorphism $\partial$ is trivial. We have $\mathrm{St}_{S_n}(s_0) = \{\text{$\sigma \in S_n$ centralising $\mathrm{Im}(\pi_1 f : \pi_0 G \to S_n)$}\}$. The discussion above then gives
\begin{align*}
H^{2}(BG ; (\Bbbk^\ast)^{\oplus n})/\mathrm{St}_{S_n}(s_0) &\overset{\sim}\to \{\text{those elements of $\pi_0({^* \TFT_{2,2}^{hG}})$ inducing the splitting $\pi_1 f$}\}\\
0 \to H^{1}(BG ; (\Bbbk^\ast)^{\oplus n}) &\to \pi_1({^* \TFT_{2,2}^{hG}}, s) \to \mathrm{St}_{S_n}(s_0) \to 0\\
H^{0}(BG ; (\Bbbk^\ast)^{\oplus n}) &\overset{\sim}\to \pi_2({^* \TFT_{2,2}^{hG}}, s).
\end{align*}

For example, taking $G=S_n$ and $f : BS_n \to X_n$ to be the map that acts on the algebra $\Bbbk^n$ by permuting the factors, then for $n \geq 3$ there are no elements centralising all of $S_n$ so 
\begin{align*}
\pi_1({^* \TFT_{2,2}^{hG}}, s) &\cong H^{1}(BS_n ; (\Bbbk^\ast)^{\oplus n}) \cong H^1(BS_{n-1} ; \Bbbk^\ast) \cong \Bbbk^\ast[2]\\
\pi_2({^* \TFT_{2,2}^{hG}}, s) &\cong H^{0}(BS_n ; (\Bbbk^\ast)^{\oplus n}) \cong \Bbbk^\ast.
\end{align*}
On the other hand if $n=2$ then $H^{1}(BS_2 ; (\Bbbk^\ast)^{\oplus 2}) \cong H^1(BS_{1} ; \Bbbk^\ast)=0$ so
\begin{align*}
\pi_1({^* \TFT_{2,2}^{hG}}, s) &\cong S_2\\
\pi_2({^* \TFT_{2,2}^{hG}}, s) &\cong H^{0}(BS_2 ; (\Bbbk^\ast)^{\oplus 2}) \cong \Bbbk^\ast.
\end{align*}
These groups are abstractly isomorphic to the above if $\Bbbk$ does
not have characteristic 2, but their origin, and presumably therefore
the physical interpretation, is different.

As another example, let $G$ be a connected group and $f : BG \to X_n$ be a map. This map must be trivial on $\pi_1$, so $\pi_0({^* \TFT_{2,2}^{hG}}) \cong H^2(BG ; (\Bbbk^\ast)^{\oplus n})/S_n$, and the $f$ corresponds to an $S_n$-orbit of an element $\xi \in H^2(BG ; (\Bbbk^\ast)^{\oplus n}) = H^2(BG ; \Bbbk^\ast)^n$. Such a theory has
\begin{align*}
\pi_1({^* \TFT_{2,2}^{hG}}, s) &\cong \{\text{stabiliser of $S_n$-action on $\xi$}\}\\
\pi_2({^* \TFT_{2,2}^{hG}}, s) &\cong H^{0}(BG ; (\Bbbk^\ast)^{\oplus n}) \cong (\Bbbk^\ast)^{\oplus n}.
\end{align*}
This generalises \cite[Lemma 3.3.1]{davidovich_2011}. In particular,
letting $G=SO(2)$ act trivially on $^* \TFT_{2,2}$ we obtain a space with $\pi_0({^* \TFT_{2,2}^{hSO(2)}}) \cong
H^2(BSO(2) ; (\Bbbk^\ast)^{\oplus n})/S_n \cong  (\Bbbk^\ast)^{\oplus
  n})/S_n$. Since the action of $SO(2) \subset O(2)$  via tangential symmetries  on $^*
\TFT_{2,2}$ trivializes \cite{davidovich_2011}, and $Spin^r(2) \cong SO(2)$, this reproduces the classification 
of TFTs with $r$-spin structure for any $r \geq 1$ \cite{schommer2009classification,Carqueville:2021cfa}, and shows that abstractly their classification is independent of $r$. Namely \cite{schommer2009classification},
since every algebra is Morita equivalent to 
$\Bbbk^n$, the components are given by a choice of $n$ and a choice of
Frobenius structure on $\Bbbk^n$. The latter is classified by the
trace ({\em i.e}. identity) map
on each factor of $\Bbbk$, each of which may be multiplied by a
non-vanishing (to ensure non-degeneracy)
element in $\Bbbk$, {\em i.e.} an element in $\Bbbk^*$, up to
permutation. However, the map ${^* \TFT_{2,2}^{hSO(2)}} \to {^*
\TFT_{2,2}^{hSpin^r(2)}}$ induced by the covering map
$Spin^r(2) \to  SO(2)$ is not an equivalence: the functoriality
with respect to $G$ of our arguments above allows us to see that on $\pi_0$ it
sends each of the $n$ elements of $\Bbbk^*$ to its $r$th power,
while it induces an isomorphism on all higher homotopy
groups. In particular, we note that if two oriented theories differ in their structure-constants by $r$th roots of unity, then they become isomorphic as $r$-spin theories.

%%%%%%%%%%%%%%%%%%%%%%%%%%%%%%%%%%%%%%
\section{Non-maximally-extended examples\label{sec:egnmext}} 
%%%%%%%%%%%%%%%%%%%%%%%%%%%%%%%%%%%%%%
Here we discuss the example of TFTs in $d=2$ based on ordinary
categories rather than bicategories.
We will see by means of an example that, even by taking just one step down the ladder
compared to the maximally-extended case, unphysical anomalies can
arise, in that a globalization
map can fail to be $\pi_0$-surjective.

We consider the well-studied case of oriented topological
field theories in $d=2$ with a finite group internal gauge
symmetry. So we consider the homomorphism $\chi:G\times SO(2)
\rightarrow O(2):(g,s) \mapsto s$, with $G$ finite, along with the
globalization map
\begin{equation}
	\Pi_*:{^{G\times SO(2)}\TFT_{2,1}}\rightarrow \Map_{G}(EG,{^{SO(2)}\TFT_{2,1}})\cong \Map(BG,{^{SO(2)}\TFT_{2,1}}).
      \end{equation}

To describe the category ${^{G\times SO(2)}TFT_{2,1}}$, let us
  introduce some definitions. A {\em Frobenius $G$-algebra over $\Bbbk$} is a pair
  $(A,\eta)$ consisting of a $G$-graded $\Bbbk$-algebra $A$ (
so $A=\oplus_{g\in G} A_g$ such that $A_gA_h\subseteq A_{gh}$), and
$\eta:A \times A\rightarrow\Bbbk$ is a $\Bbbk$-bilinear form,
or equivalently a $\Bbbk$-linear map $A\otimes  A\rightarrow\Bbbk$, such that:
$\eta(A_g\otimes A_h)=0$ if $gh\ne 1$; $\eta$ is non-degenerate when
restricted to $A_g\otimes A_{g^{-1}}$; and $\eta(ab,c)=\eta(a,bc)$~\cite{turaev2010homotopy}. A {\em Frobenius algebra} is a Frobenius
$G$-algebra with $G$ given by the trivial group.

 A \emph{crossed $G$-algebra over $\Bbbk$} is a triple $(A,\eta,\phi)$,
 where $(A,\eta)$ is a Frobenius $G$-algebra over $\Bbbk$  and $\phi:G\rightarrow
 \mathrm{Aut}(A)$ is a group homomorphism such that: $\phi_g$ preserves $\eta$; $\phi_g(A_g)\subseteq
 A_{ghg^{-1}}$; $\phi_g|_{A_g}=\mathrm{id}$; for $a \in A_g$
 and $b \in A_h$ then $\phi_h(a)b=ba$; and for $g,h\in G$ and $c \in
 A_{ghg^{-1}h^{-1}}$ we have that $\mathrm{Tr}(c\phi_h:A_a\rightarrow
A_a)=\mathrm{Tr}(\phi_{g^{-1}} c: A_b\rightarrow A_b)$~\cite{turaev2010homotopy}. We remark that a crossed $G$-algebra for $G=\ast$ is a commutative Frobenius algebra.
 
The category  ${^{G\times SO(2)}TFT_{2,1}}$ is equivalent to that
    whose objects are crossed $G$-algebras
 and whose morphisms are unital algebra maps $f:A\rightarrow B$ which are $G$-equivariant and preserve $\eta$~\cite{turaev1999homotopy}.
 In turn,  ${^{SO(2)}\TFT}_{2,1}$ is equivalent to the category  whose objects are
 commutative Frobenius algebras with unit and whose morphisms are Frobenius algebra maps.

On the right-hand side of our globalization map, we therefore
  have the category
  whose objects are commutative Frobenius algebras equipped with a
  $G$-action and whose morphisms are $G$-equivariant
  isomorphisms of commutative Frobenius algebras.

 From this point of view, the globalization map takes
 $(A,\eta)$ to its {\em principal component} $(A_e,\eta|_{A_e\otimes A_e})$
 equipped with the homomorphism which sends $g\in G$ to $\phi_{g}|_{A_e}$~\cite{turaev1999homotopy}.

To exhibit an unphysical  `t~Hooft
   anomaly, observe that for an object in ${^{SO(2)}\TFT_{2,1}^{hG}}$
   to be gaugeable, we require that for each $g \in G$ the
   corresponding algebra morphism $\phi_g$ have integer
   trace~\cite{turaev2010homotopy}. An example of an object in
   ${^{SO(2)}\TFT_{2,2}^{hG}}$ which fails this criterion may be given as follows: Let $A=\mathbb{C}[x,y]/( x^2, y^2)$ with trace $\eta(1,xy)=1$ and $\eta(1,r)=0$ for $r=1,x,y$. The automorphism 
   \begin{align*}
   	x\mapsto u x, \quad  y \mapsto u^{-1} y
   \end{align*}
   has trace given by $2+u+u^{-1}$ which is generically non-integral. Thus if the action of $G$ on $A$ involves such an automorphism it is not gaugeable.
   
  We remark that the algebra $A$ is not semi-simple, since any
  semi-simple Frobenius algebra only has integral traced
  automorphisms. Thus we know that $A$ cannot descend from a
  fully-extended theory by looping.

%%%%%%%%%%%%%%%%%%%%%%%%%%%%%%%%%%%%%%
\section{Lie group symmetries and Noether's theorem\label{sec:lie}} 
%%%%%%%%%%%%%%%%%%%%%%%%%%%%%%%%%%%%%%
Up until now, we have treated the individual $q$-form symmetry groups for each $q$ of
topological field theories as abstract groups, without the smooth
({\em i.e.} Lie group) structure with which symmetries in physics are
usually endowed. Doing so allowed us to package the tower of $q$-form
symmetries into a single topological group, and so on. But it brings
significant disadvantages. In particular, there is no possibility of
deriving a generalized version of Noether's theorem, which associates
conserved currents to a Lie group symmetry.

In this Section, we take a first step in the direction of extending
our results to Lie groups by sketching
a version of
Noether's theorem for ordinary global Lie group symmetries of unextended
oriented topological field theories.

Let us first ask roughly what form this `theorem' might take. In
physics, we have a Lie group of symmetries of a theory and Noether's
theorem gives us, for each element in the corresponding Lie algebra, a
conserved current or a conserved charge. The conserved current is
usually thought of as a vector or a 1-form, but this requires a Hodge
structure of some kind (\emph{e.g.} from a metric), which is not available to
us in topological field theory. In fact, the conserved current arises as a
differential form whose degree is one less than the dimension of spacetime $W$. Current conservation is then
simply  the
statement that the form is closed.

The degree of the form is such that one can integrate it on a
closed oriented submanifold $M$ of codimension one (or more generally a closed
cycle). We call the value of the integral, which will vanish if the
manifold bounds in $W$, the charge on $M$. This can be viewed as a
vast generalization of the usual notion in physics that `the charge is conserved',
which amounts to the statement that the charge evaluated on one connected component of the boundary of
$M\times I$ equals to the charge on the other connected component.
Importantly, it allows for spacetime evolutions that are topologically
non-trivial, which is just as well, since these are the only
non-trivial evolutions in a TFT.

In the above, we tacitly assumed  that our theory was a classical one,
in which the
conserved charge is a number (obtained from fields satisfying the
equations of motion). In quantum field theory, we instead obtain the
Ward identities for correlation functions involving the conserved
charge. This is what we want to reproduce in TFT. We shall do so in
the following way: a correlator is interpreted as the result of
applying a functor to a bordism to the empty set obtained by cutting out
tubular neighbourhoods of the supports of the operators appearing in
the correlator.  For a conserved charge, this means a submanifold $M$ of codimension one, whose normal bundle
will be trivial (since everything is oriented) and whose tubular
neighbourhood has boundary $M \coprod \overline{M}$. We will see that we
get a map $\mathfrak{g} \to Z(M) \otimes Z(M)^\vee$ which picks out the
conserved charges. They are conserved in the sense that, if the
bordism when cut contains a piece $M \to \emptyset$, then the correlation
function vanishes when evaluated on elements of $Z(M) \otimes
Z(M)^\vee$ in the image of the above map from $\mathfrak{g}$. The
requirement that the bordism factors in this way corresponds
to our earlier requirement that the conserved charge is to be evaluated on a
cycle that bounds.

Let us see in more detail how this happens. We suppose that we are
given a TFT whose automorphism group can be given a smooth
structure, such that it acts smoothly on the vector space $Z(M)$ assigned to
each object $M$ in the bordism category. For example, for oriented theories in $d=1$, we have seen that this group is
$GL(n,\Bbbk)$ for some $n$, so this is certainly the case for $\bbR$
or $\bbC$. We take a Lie group $G$ and a smooth homomorphism into the
automorphism group (which corresponds to our earlier notion of the
trivial action of the group on the space of TFTs).

Differentiating the map $G \to GL(Z(M))$ for each object then gives a
map $\mathfrak{g} \to
\mathrm{End} (Z(M))$, which, since $Z(M)$ is finite-dimensional, is
canonically isomorphic to $Z(M)\otimes Z(M)^\vee$.
The latter space can be interpreted as the vector space of operators associated to a
codimension one submanifold $M$ of some spacetime $W$. Indeed, when
everything is oriented, the normal bundle of $M$ in $W$ is necessarily
trivial and so the boundary of a tubular neighbourhood is given by two
copies of $M$ with opposite orientations, to which the TFT functor
assigns $Z(M)\otimes Z(M)^\vee$.

Now let us consider the effect of inserting these operators into
correlation functions.
Let $W:N\rightarrow N^\prime$ be a cobordism with
$M$ embedded in the interior of $W$.
Excising an open tubular neighbourhood of $M$ in $W$, we obtain
another cobordism $\tilde W:N\coprod M\coprod M^\vee\rightarrow
N^\prime$. Given $\xi \in \mathrm{End} Z(M)$, we form the linear map 
\begin{align} \label{eq:Correlator}
	Z(N)\xrightarrow{\mathrm{id}\otimes \tilde \xi} Z(N)\otimes Z(M)\otimes Z(M^\vee)\xrightarrow{Z(\tilde W)} Z(N^\prime)
\end{align}
where $\tilde \xi: \Bbbk \rightarrow  Z(M)\otimes Z(M^\vee)$ is the
linear map such that $\tilde \xi (1)$ is the element corresponding to
$\xi$.
Our map $\mathfrak{g} \to
Z(M)\otimes Z(M)^\vee$ picks out a distinguished subspace of such
linear maps and gives it the structure of an algebra, just as we
expect for conserved charges.

To see the sense in which charges are conserved, force
$N^\prime$ to be the empty set. Combining $\mathfrak{g}\rightarrow
\mathrm{End}(Z(M))$ and (\ref{eq:Correlator}) we get a natural
map $Z(N)\otimes\mathfrak{g}\rightarrow\Bbbk$, but it is easy to show
that this map is in fact the zero map. Translating back to physics, we
see that any correlation function with an insertion of a conserved
charge on $M$ vanishes, when $M$ is nullbordant in $W$, as desired.

%%%%%%%%%%%%%%%%%%%%%%%%%%%%%%%%%%%%%%
\section{Discussion\label{sec:disc}}
%%%%%%%%%%%%%%%%%%%%%%%%%%%%%%%%%%%%%%
In closing, we would like to show how our results shed some light on
the 
previous literature. We begin with `t~Hooft anomalies of ordinary
symmetries of non-maximally
extended
orientable TFTs, which
were studied in ~\cite{Kapustin:2014zva} (and formalized and generalized
in~\cite{Muller:2018doa,Muller:2020phm}). In our notation (omitting the $SO(2)$ factors
corresponding to orientation), the idea
can be described as follows. Letting $K$ be an ordinary group acting
trivially on $\TFT_{2,1}$,
a TFT in $\TFT_{2,1}^K \simeq \Map(BK,\TFT_{2,1})$ may not have a preimage in
$^K\TFT_{2,1}$ under the obvious globalization map. In~\cite{Kapustin:2014zva}, this problem was studied
using an approach which is natural from the physicist's (if not the
physics) point of view, namely to study $K$ symmetries of quantum
theories which arise by
quantizing classical theories. We
have not discussed quantization in the present work, but for the
discussion that follows it suffices to know that there exists an
{\em orbifoldization functor}~\cite{schweigert2019orbifold} from the category of finite groups to
the category of spaces that sends a group $G$ to $^G \TFT$ and that for
the homomorphism $G \to \ast$ to the trivial group this corresponds,
when restricted to invertible theories in $^G \TFT$ to
quantization \`{a} la Dijkgraaf--Witten~\cite{Dijkgraaf:1989pz}.

Given a short exact sequence $\ast \to H \to G \to K \to \ast$ of
finite groups, there is a commutative diagram
\begin{equation}\label{eq:Mueller}
		\begin{tikzcd}
		 {^G\TFT_{2,1}}\arrow{r}\arrow[swap]{d}{\mathrm{Glob}}&{^K\TFT_{2,1}} \arrow{d}{\mathrm{Glob}}\\
			\Map_K(EK, {^H\TFT_{2,1}})\arrow{r}& \Map(BK,\TFT_{2,1})
		\end{tikzcd}
              \end{equation}
The proposal of ~\cite{Kapustin:2014zva} is, given $K$, to start from an invertible theory in
$^H\TFT_{2,1}$ (with the trivial $H$ action) and to ask if there exists an extension $G$ of $K$ by
$H$ and a theory in the resulting ${^G\TFT_{2,1}}$ (with the trivial $G$ action)
that maps to it under the left-hand map followed by composition with
the map that forgets the $K$ symmetry. If so, one may say that the
corresponding theory in $\Map_K(EK, {^H\TFT_{2,1}})$ is non-anomalous,
since one can follow the diagram to find a theory in the top
right-hand corner that is a gauging of the theory in the bottom right-hand
corner that is the quantization of the  theory in $\Map_K(EK,
{^H\TFT_{2,1}})$. In ~\cite{Kapustin:2014zva} and
~\cite{Muller:2018doa,Muller:2020phm}, obstructions to this were given,
as well as sufficient conditions for the construction to go through.

Though this construction is well-motivated from the physics point of view, it
seems unreasonable to us to describe failures of this construction
as anomalies. When the construction fails, one cannot even
find a quantum theory with global $K$ symmetry in
$\Map(BK,\TFT_{2,1})$ for which one can ask the question of whether $K$
is gaugeable. Moreover, any theory in $K$ which one actually can
construct by quantizing a classical theory is automatically gaugeable, because the
Dijkgraaf--Witten construction extends to fully-extended theories~\cite{morton2015cohomological},
where the cobordism hypothesis applies.

The same trick of extending theories can be applied to understanding
the results of~\cite{Gukov:2021swm}, where it is shown that $0$- and $1$-form
symmetries of a sub-class of oriented non-maximally extended TFTs
in $d=2$ ({\em i.e.} $^{SO(2)}\TFT_{2,1}$) valued in $\Vect_\bbC$, to wit those corresponding to commutative Frobenius algebras
that are additionally semisimple,
are given by permutations and phasings, respectively. These are the
symmetries in the intrinsic sense of being automorphisms of the TFTs
and the result comes as no surprise once we observe that the
semisimple algebras are the ones that arise from fully-extended TFTs
upon looping. So the automorphism groups can be read off directly from
(\ref{eq:DavidovichEq})
and correspond to permutations of the simple factors that preserve the
trace map and rephasings (or rather elements in $\bbC^\ast$ given
that we have no inner product structure).

Finally, generalized symmetries of certain extended, oriented
topological field theories in $d=2$ are described in \cite{Gaiotto:2014kfa}, namely
those obtained by quantizing classical $\mathbb{Z}_{n_1}\times
\mathbb{Z}_{n_2}$ gauge theories. The resulting algebra is the twisted
group algebra $\mathbb{C}_{\omega_p}[\mathbb{Z}_{n_1}\times
\mathbb{Z}_{n_2}]$ where $\omega_p((a_1,b_1),(a_2,b_2))=\exp(2\pi \mathrm{i} p
a_1 b_2/\mathrm{GCD}(n_1,n_2))$. The algebra  $\mathbb{C}_{\omega_p}[\mathbb{Z}_{n_1}\times
\mathbb{Z}_{n_2}]$  is Morita equivalent (and therefore equivalent in  $\TFT_{2,2}$) to the algebra $\Bbbk^{n_1n_2/m^2}$ where $m=\mathrm{GCD}(n_1,n_2)/\mathrm{GCD}(p,n_1,n_2)$ is the dimension of the irreducible projective representation with twisting $\omega_p$, and $n_1n_2/m^2$ their number. Correspondingly the automorphism groups are $S_{n_1n_2/m^2}$ and $(\bbC^\ast)^{n_1n_2/m^2}$, which are substantially larger than the $\mathbb{Z}_{n_1/m}\times
\mathbb{Z}_{n_2/m}$ subgroups obtained in \cite{Gaiotto:2014kfa} by inspection of the classical action.
This example shows that, at least for spaces of topological field
theories with their large number of equivalences, studying classical
actions gives a rather poor guide to the resulting quantum
physics. Happily, the power of the cobordism hypothesis for physical
({\em i.e.} fully-extended) theories suggests that we may one day no longer need
to.
%%%%%%%%%%%%%%%%%%%%%%%%%%%%%%%%%%%%%% 
\section{Acknowledgments\label{sec:ack}} 
%%%%%%%%%%%%%%%%%%%%%%%%%%%%%%%%%%%%%%
BG is supported by STFC consolidated grant ST/T000694/1. JTS is supported in part by the U.S. National Science Foundation (NSF) grant PHY-2014071.
    %%%%%%%%%%%%%%%%%%%%%%%%%%%%%%%%%%%%%%%%%%%%%%%%%%%%%%%%%%%%%%%%%
\bibliography{generalized_symmetry}

  %%%%%%%%%%%%%%%%%%%%
 \end{document}